\documentclass[fleqn,usenatbib]{mnras}

\usepackage{newtxtext,newtxmath}

\usepackage[T1]{fontenc}

\usepackage{graphicx}	
\usepackage{amsmath}	
\usepackage{amssymb}	
\usepackage{multirow}
\usepackage{bm}

\newcommand{\Gpc}{\ifmmode  {\rm~Gpc}  \else ${\rm~Gpc}$\fi}
\newcommand{\Mpc}{\ifmmode  {\rm~Mpc}  \else ${\rm~Mpc}$\fi}
\newcommand{\kpc}{\ifmmode  {\rm~kpc}  \else ${\rm~kpc}$\fi}
\newcommand{\Msun}{\ifmmode {\rm M}_{\odot} \else ${\rm M}_{\odot}$ \fi} 
\newcommand{\Msunpyr}{\ifmmode M_{\odot}{\rm~yr}^{-1} \else $M_{\odot}{\rm~yr}^{-1}$ \fi}

\newlength{\fullwidth}
\newlength{\halfwidth}
\title[GalaxyNet: deep learning galaxy formation]
{GalaxyNet: Connecting galaxies and dark matter haloes with deep neural networks and reinforcement learning in large volumes}

\author[B. P. Moster et al.]{
Benjamin P. Moster,$^{1,2}$\thanks{E-mail: moster@usm.lmu.de (BPM)}
Thorsten Naab,$^{2}$
Magnus Lindstr\"om,$^{2}$
Joseph A. O'Leary$^{1}$
\\
$^{1}$Universit\"ats-Sternwarte, Ludwig-Maximilians-Universit\"at M\"unchen, Scheinerstr. 1, 81679 M\"unchen, Germany\\
$^{2}$Max-Planck-Institut f\"ur Astrophysik, Karl-Schwarzschild Stra\ss e 1, 85748 Garching, Germany\\
}

\date{Accepted XXX. Received YYY; in original form ZZZ}

\pubyear{2020}

\begin{document}
\label{firstpage}
\pagerange{\pageref{firstpage}--\pageref{lastpage}}
\maketitle

\begin{abstract}
We present the novel wide \& deep neural network GalaxyNet, which connects the properties of galaxies and dark matter haloes, and is directly trained on observed galaxy statistics using reinforcement learning. The most important halo properties to predict stellar mass and star formation rate (SFR) are halo mass, growth rate, and scale factor at the time the mass peaks, which results from a feature importance analysis with random forests. We train different models with supervised learning to find the optimal network architecture. GalaxyNet is then trained with a reinforcement learning approach: for a fixed set of weights and biases, we compute the galaxy properties for all haloes and then derive mock statistics (stellar mass functions, cosmic and specific SFRs, quenched fractions, and clustering). Comparing these statistics to observations we get the model loss, which is minimised with particle swarm optimisation. GalaxyNet reproduces the observed data very accurately ($\chi_\mathrm{red}=1.05$), and predicts a stellar-to-halo mass relation with a lower normalisation and shallower low-mass slope at high redshift than empirical models. We find that at low mass, the galaxies with the highest SFRs are satellites, although most satellites are quenched. The normalisation of the instantaneous conversion efficiency increases with redshift, but stays constant above $z\gtrsim0.7$. Finally, we use GalaxyNet to populate a cosmic volume of $(5.9\Gpc)^3$ with galaxies and predict the BAO signal, the bias, and the clustering of active and passive galaxies up to $z=4$, which can be tested with next-generation surveys, such as LSST and Euclid.
\end{abstract}





\section{Introduction}
\label{sec:intro}

The connection between galaxies and dark matter haloes is of fundamental importance to astrophysics and cosmology. In the standard $\Lambda$CDM framework, structure formation proceeds through gravitationally driven hierarchical collapse and merging. From a relatively smooth initial state, the matter distribution of the Universe evolved into a complex cosmic web, in which small perturbations developed into extended gravitationally bound objects called haloes. Within this picture, galaxies form by the cooling and condensation of gas in the centres of virialised dark matter haloes \citep{Rees:1977aa,White:1978aa,Fall:1980aa,Blumenthal:1984aa}. Consequently, the properties of galaxies strongly depend on the properties of the dark matter haloes in which they form. 

The formation of dark matter haloes has been studied extensively with large cosmological $N$-body simulations \citep[e.g.][]{Springel:2005ab,Klypin:2016aa}, which has given us a fairly detailed picture of their abundance, clustering, and substructure. The derived halo mass function (HMF) is very steep over many orders of magnitude. Massive haloes are clustered more strongly compared to their lower-mass counterparts, and haloes that assemble early are more strongly clustered at fixed final mass \citep[halo assembly bias;][]{Gao:2005aa}. With the advent of deep surveys, the properties of galaxies can now also be determined up to high redshift. The stellar mass function (SMF) increases from high to low redshift, and has an exponential cut-off above a characteristic mass \citep[e.g.][]{Li:2009aa,Ilbert:2013aa,Muzzin:2013aa}. Like haloes, galaxies are biased tracers of the density field, with an increasing bias for more massive galaxies and higher redshift, and a larger bias for passive galaxies at fixed stellar mass \citep[e.g.][]{Guo:2011aa}. If the conversion of gas into stars were equally efficient in all haloes, we would expect the SMF to have the same shape as the HMF. The different shapes therefore indicate the complexity of the baryonic physics regulating galaxy formation. Moreover, if the properties of galaxies were only depending on halo mass, we would expect the same clustering properties for active and passive galaxies. The observed galaxy assembly bias therefore indicates that galaxy properties also depend on secondary halo properties, such as the age.

Several different approaches have been used in the past to link galaxy properties to the properties of their dark matter haloes. One option is to infer the mass of individual dark matter haloes from the observed galaxy properties, for example though galaxy kinematics \citep{More:2009aa,Wojtak:2013aa,Lange:2019aa}, gravitational lensing \citep{Mandelbaum:2006aa,Mandelbaum:2016aa,Hudson:2015aa} or the properties of X-ray haloes \citep{Kravtsov:2018aa}. Another possible approach are models that employ a specified set of physical processes for the baryons to evolve an initial distribution of dark matter and gas, such as \textit{hydrodynamic simulations} \citep{Springel:2005ac,Teyssier:2013aa,Hopkins:2014aa,Wang:2015aa} and \textit{semi-analytic models} \citep{White:1991aa,Kauffmann:1993aa,Cole:1994aa,Somerville:1999aa}. While this methods aims to follow the physical processes, in practice the limited achievable resolution requires the use of simplified recipes or sub-grid models, and the adjustment of the free parameters using observational  constraints.

An alternative option avoids explicitly modelling the baryonic physics, but relates the observed galaxy populations to the underlying dark matter haloes in a statistical manner that is as independent as possible of any model assumptions on poorly understood baryonic physics, `marginalising' over these uncertainties. To this end, these \textit{empirical models} of galaxy formation adopt parameterised relations between galaxy and halo properties. The models are trained, i.e. the parameters are determined, by requiring a set of observations be reproduced. One of the most simple models, \textit{subhalo abundance matching} assumes that each halo hosts one central galaxy and each subhalo hosts one satellite galaxy, while there is a simple monotonic relation between stellar and halo mass such that the SMF is recovered (\citealt{Vale:2004aa}; \citealt*{Conroy:2006aa}; \citealt{Moster:2010aa,Moster:2013aa}; \citealt{Behroozi:2010aa,Behroozi:2013aa}; \citealt{Grylls:2020aa}). State-of-the-art empirical models are trained on several data sets and are now able to reproduce the galaxy assembly bias and colour-dependent clustering as they rely on secondary halo properties beyond halo mass, which is achieved either by matching galaxies and haloes by age \citep{Hearin:2013aa}, or by self-consistently following the formation history of haloes and galaxies \citep{Moster:2018aa,Behroozi:2019aa}.

An important advantage of empirical models is their ability to make predictions of galaxy properties unbiased by assumptions on poorly understood baryonic physics \citep[e.g.][]{Hearin:2019aa,Moster:2019aa,OLeary:2020aa}. However, the difficulty for empirical models is to find sensible relations between galaxy and halo properties. Typically, the parameterised functions that are used for these relations are chosen a priori, and model selection techniques are employed to find the one that is best suited to reproduce the data. This process can be very tedious, and it is not guaranteed that one finds the relations that lead to the best possible agreement with observations. A far improved approach is to use algorithms that can automatically connect galaxies and haloes without the need to specify a parameterised relation.

In the last decade, the field of \textit{artificial intelligence} (AI) and \textit{machine learning} (ML) has vastly expanded, and several ML methods have recently been used in astrophysics. Their big advantage is that they give computers the ability to learn from data without being explicitly programmed. Whereas for classical numerical methods, such as hydrodynamic simulations and current empirical models we need to know all (complex) `rules' beforehand, an ML algorithm can detect patterns automatically. In general, an ML algorithm processes the input values, called \textit{features}, and maps them to some output values, called \textit{targets}. While there are many different types of ML systems, they can be broadly categorised on whether or not they are trained with supervision. In \textit{supervised learning}, the training data includes the desired target solution (called \textit{labels}). To train a system, it needs to be given many examples, including their features and labels. Typical tasks are regression and classification. In \textit{unsupervised learning} the training data is unlabeled, and the system learns without supervision, e.g. clustering and dimensionality reduction algorithms. Finally, \textit{reinforcement learning} does not need labels for each data instance. Instead the system is rewarded or penalised at specific points during learning, based on how it performs. It must then learn by itself what is the best strategy to optimise a global task (e.g. earn the most reward over time). This approach has been taken to teach robots how to walk, or to train machines how to play games \citep[e.g. AlphaGo;][]{Silver:2016aa,Silver:2017aa}.

Among the most popular supervised learning algorithms are \textit{decision trees} \citep{Breiman:1984aa}, in which leaves represent target values and branches represent conjunctions of features that lead to those target values, and \textit{random forests} \citep{Breiman:2001aa}, which construct a multitude of decision trees at training time and output the target value that is the mode of the individual trees, which corrects for overfitting. \textit{Artificial neural networks} are based on a collection of connected nodes called artificial neurons organised in layers, loosely modelling the neurons in a biological brain \citep{McCulloch:1943aa}. Each connection can transmit a signal from a neuron on one layer to a neuron on a subsequent layer. A neuron that receives input signals processes the sum of its inputs with a non-linear \textit{activation function}, and then transmits it to each connected neuron. Each connection has an associated weight that increases or decreases the strength of the signal. The weights get adjusted as learning proceeds. If the network consists of many (hidden) layers, the algorithm is commonly called `deep learning'.

In astrophysics, the number of studies that apply ML techniques has risen substantially in the last years. Unsupervised learning algorithms have been used to identify different kinematic components of simulated galaxies \citep{Obreja:2018aa,Obreja:2019aa}, to compare stellar spectra \citep{Traven:2017aa}, to classify pulsars \citep{Lee:2012aa}, and to find high-redshift quasars \citep{Polsterer:2013aa}. Supervised learning has been used to classify variable stars \citep{Richards:2011aa}, to classify galaxies morphologically \citep{Huertas:2008aa}, and to determine the redshift of galaxies \citep{Hoyle:2015aa,Hoyle:2016aa,DIsanto:2018aa}. Recently, ML has also been used to connect the properties of galaxies and dark matter haloes using supervised learning techniques. \citet{Sullivan:2018aa} train a simple neural network with one hidden layer to predict the baryon fraction within a dark matter halo at high redshift, given several halo properties (features). As training data they use the results of a cosmological hydrodynamic simulation with Ramses-RT. Similarly, \citet{Agarwal:2018aa} use several ML methods to link input halo properties to galaxy properties, training on the Mufasa cosmological hydrodynamical simulation. The limitation of both studies is the supervised training and the training data. As labelled galaxy-halo data is not available for observed systems, the data for supervised learning has to be taken from a model. Even if the ML algorithms learn to reproduce the training data perfectly, the connection between galaxy and halo properties is the same as in the simulations. If the simulations predict the true relations poorly, so will the ML method. Therefore ML algorithms should not be trained on simulated data, but on observed data directly.

The main task of this paper is to employ ML methods and automatically find the connection between observed galaxy properties and simulated dark matter haloes without imposing any relations a priori. Instead of using supervised learning and training the algorithms with data that have been obtained from another model, our aim is to directly train the algorithm with observed data using a reinforcement learning approach. To this end, we do not determine the goodness of the fit, i.e. the model loss, based on the difference between ML predictions and training labels from another model, but rather on the difference between ML predictions and observed galaxy statistics, such as the SMF, local clustering, cosmic and specific SFRs, and quenched fractions. Since we cannot compute this loss for individual galaxy-halo instances, we first calculate the properties of the total galaxy population with a wide \& deep neural network, and then reward or penalise it based on its performance on the observed statistics. We train the parameters of our network GalaxyNet (weights and biases) with a particle swarm optimisation (PSO) technique using dedicated NVIDIA tensor cores. Finally, we use the trained GalaxyNet to calculate the baryon conversion efficiency and to predict galaxy clustering for cosmological surveys.

This paper is organised as follows. In Section \ref{sec:featureselection} we first describe how features (halo properties) and labels for supervised learning (galaxy properties from \textsc{emerge}) are obtained, and then determine the feature importance using random forests, selecting the most important halo properties. In section \ref{sec:supervised}, we use supervised learning with data from \textsc{emerge} to find the optimal neural network architecture. We train GalaxyNet with reinforcement learning using the PSO in section \ref{sec:reinforcement}, and show how the galaxy properties compare to empirical models. In section \ref{sec:efficiency}, we derive the baryon conversion efficiency from GalaxyNet, and predict galaxy clustering in section \ref{sec:lss}. We conclude and discuss our results in section \ref{sec:conclusions}.

Throughout this work we assume a Planck $\Lambda{\rm CDM}$ cosmology with ($\Omega_\mathrm{m}$, $\Omega_\mathrm{\Lambda}$, $\Omega_\mathrm{b}$, $h$, $n_\mathrm{s}$, $\sigma_8$) = (0.307, 0.693, 0.0485, 0.6777, 0.9611, 0.8288). We employ a \citet{Chabrier:2003aa} initial mass function (IMF) and convert all stellar masses and SFRs to this IMF. All virial masses are computed according to the overdensity criterion by \citet{Bryan:1998aa}. To simplify the notation, we will use the upper case $M$ to denote dark matter halo masses and the lower case $m$ to denote galaxy stellar masses.


\section{Feature Selection}
\label{sec:featureselection}

The first step to learn the connection between galaxies and haloes is to select the features, i.e. halo properties, that are used to derive the targets, i.e. galaxy properties. While some features will be very important to get an accurate prediction for the targets, some other features may have little to no impact on the targets and can therefore safely be neglected. This reduces the complexity of the algorithm and therefore helps to prevent overfitting. To select the most appropriate features, we perform a feature importance analysis using random forests. For this task, we need features which we extract from a large $N$-body simulation, and target labels which we calculate with the empirical galaxy formation model \textsc{emerge}. Both features and labels are scaled before processed with a random forest regressor which determines the feature importance based on how often a feature is used in the forest.


\subsection{Features: Properties of Simulated Haloes}
\label{sec:features}

The features in this ML task are the properties of dark matter haloes which we extract from a numerical simulation with $200\Mpc$ side length adopting a cosmology consistent with the latest results by the \citet{Planck-Collaboration:2018aa}. We employed the \texttt{CAMB} code \citep*{Lewis:2000aa} to compute the initial power spectrum, and the {\sc Music} code \citep{Hahn:2011aa} to generate the initial conditions for the simulation, which contains $1024^3$ collisionless particles with a mass of $2.92\times10^8\Msun$. The TreePM code {\sc Gadget3} \citep{Springel:2005aa} was used to run the simulation with periodic boundary conditions and a gravitational softening of $3.3\kpc$ from redshift $z=63$ to 0, saving 94 snapshots equally spaced in scale factor ($\Delta a=0.01$). We identified the dark matter haloes and subhaloes in each snapshot with the halo finder {\sc Rockstar} \citep*{Behroozi:2013ac} using the criterion by \citet{Bryan:1998aa} to derive halo masses. With a minimal particle number of 100 for each halo the minimally resolved halo mass is $\log_{10}(M_\mathrm{min}/\Msun)=10.5$. We generated halo merger trees with the {\sc ConsistentTrees} code \citep{Behroozi:2013ad}. 

Due to the finite mass resolution, subhaloes can no longer be identified once tidally stripped below the resolution limit. This mass loss can be substantial, so that a special treatment becomes necessary for the haloes associated with `orphan galaxies'. To this end, we determine the orbital parameters at the last moment a subhalo is identified and apply the dynamical friction estimate by \citet*{Boylan-Kolchin:2008aa}. The disrupted subhalo is kept until the dynamical friction time has elapsed. While orbiting, the distance between subhalo and host halo centre decays and the subhalo mass declines exponentially at the same rate since reaching its peak mass. If its host halo merges with a larger halo, we recompute the dynamical friction time with respect to the new host halo, and let the subhalo merge with it once this new time has elapsed.

From these halo merger trees we select all haloes at 10 different snapshots corresponding to the redshifts $z =$ 0, 0.1, 0.2, 0.5, 1.0, 2.0, 3.0, 4.0, 6.0, and 8.0. For each halo we extract 10 properties including current properties such as halo mass and scale factor, as well as historic properties such as halo peak mass and the growth rate at the time the peak mass is reached. All extracted halo properties with a short description are listed in Table \ref{tab:features}. The growth rates are computed over one dynamical time. The growth rate $(\mathrm{d} M / \mathrm{dt})_\mathrm{p}$ is taken at the time $t_\mathrm{p}$, when the halo reaches its peak mass, and does not correspond to the maximal growth rate during the formation history. The half-mass scale factor $a_{1/2}$ corresponds to the time a halo reached half its peak mass for the first time, i.e. before the peak mass was reached. The term `main halo' is used to refer to distinct haloes that are not located within a larger halo, while all other haloes are called `subhaloes'. Throughout this work we assume that a `central galaxy' is located at the centre of a main halo, and a `satellite galaxy' within each subhalo.


\begin{table}
	\centering
	\caption{The halo properties extracted from simulations that are used as features in the ML algorithms. The second column indicates if the logarithm of the feature is taken before scaling. The third and fourth column give the minimal and maximal value for the scaling.}
	\label{tab:features}
	\begin{tabular}{ccccl}
		\hline
		Feature & $\log$ & Min & Max & Description\\
		\hline
		$M_\mathrm{h}$ & Y & $10^{10}$ & $10^{16}$ & Current virial mass in \Msun\\
		$M_\mathrm{p}$ & Y &  $10^{10}$ & $10^{16}$ & Peak virial mass (at $t_\mathrm{p}$) in \Msun\\
		$(\mathrm{d} M / \mathrm{dt})_\mathrm{h}$ & Y &  $10^{-2}$ & $10^{6}$ & Current growth rate in \Msunpyr\\
		$(\mathrm{d} M / \mathrm{dt})_\mathrm{p}$ & Y & $10^{-2}$ & $10^{6}$ & Growth rate at $t_\mathrm{p}$ in \Msunpyr\\
		$a$ & N & 0.0 & 1.0 & Current scale factor\\
		$a_\mathrm{p}$ & N & 0.0 & 1.0 & Scale factor at peak mass\\
		$a_{1/2}$ & N & 0.0 & 1.0 & Scale factor at half  peak mass\\
		$R_\mathrm{h}$ & Y & 0.01 & 3.0 & Current vrial radius in \Mpc\\
		$c$ & Y & $10^{-2}$ & $10^{4}$ & Current concentration parameter\\
		$\lambda$ & Y & $10^{-3}$ & $1$ & Current spin parameter\\
		\hline
	\end{tabular}
\end{table}


\subsection{Labels: Properties of Galaxies computed by \textsc{emerge}}
\label{sec:labels}

To perform a feature importance analysis with random forests, we need to know the target value for each halo, i.e. we have to use labelled data. The targets we aim to compute in this paper are the stellar mass of a galaxy $m$, and its SFR $\dot m$. To create labels for those targets we use the empirical galaxy formation model \textsc{emerge} \citep{Moster:2018aa}, which follows the formation of dark matter haloes to derive the associated galaxy properties. The SFR of the galaxy at the centre of a dark matter halo is set to the product of the halo's baryonic growth rate and the instantaneous baryon conversion efficiency $\epsilon$:
\begin{equation} \label{eqn:sfrcen}
\frac{{\rm d}m_*}{{\rm d}t} (M,z) = \frac{{\rm d}m_\mathrm{b}}{{\rm d}t} \cdot \, \epsilon(M,z) = f_\mathrm{b} \frac{{\rm d}M}{{\rm d}t} \cdot \, \epsilon(M,z)\; ,
\end{equation}
where $f_\mathrm{b} = \Omega_\mathrm{b}/\Omega_\mathrm{m}$ is the universal baryonic fraction. The conversion efficiency $\epsilon$ describes how efficiently the gas that falls into the halo is converted into stars, and thus combines the effects of cooling, star formation, and various feedback processes. Having tested several parametrisations assessed with various model selection criteria, the efficiency is parameterised as a double power law with normalisation $\epsilon_\mathrm{N}$, characteristic halo mass $M_1$, and slopes $\beta$ and $\gamma$:
\begin{equation} \label{eqn:epsilon}
\epsilon(M,z) = 2 \;\epsilon_\mathrm{N}(z) \left[ \left(\frac{M}{M_1(z)}\right)^{-\beta(z)} + \left(\frac{M}{M_1(z)}\right)^{\gamma(z)}\right]^{-1} \; ,
\end{equation}
where $\epsilon_\mathrm{N}$, $M_1$, and $\beta$ depend linearly on the scale factor $a$, and $\gamma$ is a constant:
\begin{align}
\log_{10} M_1(z)& = M_0 + M_\mathrm{z}(1-a) = M_0 + M_\mathrm{z}\frac{z}{z+1} \; ,\\
\epsilon_\mathrm{N}(z)& = \epsilon_0 + \epsilon_\mathrm{z}(1-a) = \epsilon_0 + \epsilon_\mathrm{z}\frac{z}{z+1} \; ,\\
\beta(z)& = \beta_0 + \beta_\mathrm{z}(1-a) = \beta_0 + \beta_\mathrm{z}\frac{z}{z+1} \; ,\\
\gamma(z)& = \gamma_0 \; .
\end{align}
The star formation history of each galaxy is then a consequence of the specific path a halo has taken through the halo mass-redshift plane. For a growth rate $\dot M$ and efficiency $\epsilon(M,z)$, \textsc{emerge} can compute the SFR of the central galaxy $\dot m_*(M,\dot M,z)$ using eqn. ($\ref{eqn:sfrcen}$). Integrating this rate over cosmic time, while taking into account the fraction of mass that is being lost as a consequence of dying stars, yields the stellar mass formed in-situ.

When a halo is accreted by a larger halo, its mass begins to decline as a result of tidal stripping, and the infall of gas onto its galaxy stops, quenching the galaxy. The SFR of the galaxy is kept constant for a time $\tau$ after the halo stops growing, and is then set to zero. The quenching time is parameterised with respect to the halo's dynamical time and is longer for low mass satellites:
\begin{equation} \label{eqn:satquenching}
\tau = t_\mathrm{dyn} \cdot \tau_0 \cdot \max \left[ \left(\frac{m_*}{10^{10}\Msun}\right)^{-\tau_\mathrm{s}},1\right]  \; ,
\end{equation}
with the free parameters $\tau_0$ and $\tau_\mathrm{s}$. If a subhalo has lost enough of its mass through tidal stripping, its gravitational potential is no longer able to protect the stars in its centre. All stars are moved to the stellar halo once the mass of the subhalo has dropped below a fraction $f_\mathrm{s}$ of the peak mass:
\begin{equation} \label{eqn:satstripping}
M < f_\mathrm{s} \cdot M_\mathrm{peak} \; ,
\end{equation}
with the stripping parameter $f_\mathrm{s}$. Finally, once a subhalo has lost all of its orbital energy due to dynamical friction, the satellite galaxy will merge with the central galaxy. At this point we let a fraction of satellite stars $f_\mathrm{esc}$ escape to the halo as diffuse stellar material, resulting in a  remnant mass of
\begin{equation} \label{eqn:satmerging}
m_\mathrm{rem} = m_\mathrm{cen} + m_\mathrm{sat} \cdot (1-f_\mathrm{esc}) \; ,
\end{equation}
with the escape fraction $f_\mathrm{esc}$. All model parameters are constrained by comparing model mock observations to observed data, including SMFs, quenched fractions, cosmic and specific SFRs, and clustering. The parameter space was explored using parallel tempering \citep{Swendsen:1986aa,Geyer:1991aa} to find the most likely parameters and their credibility intervals, which are presented in Table \ref{tab:bestfit}. For more details on \textsc{emerge}, we refer the reader to the code paper \citep{Moster:2018aa} and the user-guide on the \textsc{emerge} website\footnote{http://www.usm.lmu.de/emerge}.


\begin{table}
	\centering
	\caption{The best fit model parameters for \textsc{emerge}.}
	\label{tab:bestfit}
	\begin{tabular}{lccr}
		\hline
		Parameter & Best-fit  & Upper $1\sigma$ & lower $1\sigma$ \\
		\hline
		\hline
		$M_0$ & 11.34829  & +0.03925 & -0.04153\\
		$M_\mathrm{z}$ & 0.654238  & +0.08005 & -0.07242\\
		$\epsilon_0$ & 0.009010  & +0.00657 & -0.00451\\
		$\epsilon_\mathrm{z}$ & 0.596666  & +0.02880 & -0.02366\\
		$\beta_0$ & 3.094621  & +0.15251 & -0.14964\\
		$\beta_\mathrm{z}$ & -2.019841 & +0.22206 & -0.20921\\
		$\gamma_0$ & 1.107304  & +0.05880 & -0.05280\\
		\hline
		$f_\mathrm{esc}$ & 0.562183  & +0.02840 & -0.03160\\
		$f_\mathrm{s}$ & 0.004015  & +0.00209 & -0.00141\\
		$\tau_{0}$ & 4.461039  & +0.42511 & -0.40187\\
		$\tau_\mathrm{s}$ & 0.346817  & +0.04501 & -0.04265\\
		\hline
	\end{tabular}
\end{table}


\subsection{Data Preparation and Scaling of Features and Labels}

Having defined the features as the properties of a halo, and the targets as the properties of its galaxy, we can now proceed to investigate which features are important to obtain the desired targets. However, before we feed the features to a machine learning algorithm, it is helpful to scale the features and the targets. While this is generally not required for random forests, we will later also use the features with neural networks, which tend to perform poorly with unscaled features. To this end, we apply two transformations. The first is computing the logarithmic values of some features that are distributed logarithmically, such as halo masses, growth rates, radii, concentrations and spins. The second column of Table \ref{tab:features} lists which features have been transformed to a logarithmic scale. The second transformation is normalisation using a standard min-max scaling. For each feature we define a minimum and a maximum value as given in Table \ref{tab:features}, and then apply a linear transformation that maps the minimum to 0.01 and the maximum to 1. If a feature is smaller/larger than the minimum/maximum the values are clipped, i.e. the minimum/maximum value is used.

We also scale our target labels we obtained from \textsc{emerge}, i.e. the stellar mass of the galaxy $m_*$, and its SFR $\mathrm{d} m_* / \mathrm{dt}$. We take the logarithm of both, and use a min-max scaling with minimum and maximum values of $10^{6}$ and $10^{13}\Msun$ for $m_*$ and $10^{-6}$ and $10^4\Msunpyr$ for the SFR. To get predictions from the machine learning algorithms, we unscale the outputs using the inverse transformations. To be able to assess the hyperparameters of the machine learning algorithms and the final performance, we divide the data consisting of $\sim5,000,000$ haloes randomly into training (80 per cent), validation (10 per cent) and test set (10 per cent). In the following the training set is then purely used to train the models, the validation set is used to asses how well each model performs for selected hyperparameters, and the test set is only used to determine the final performance of the best model.


\subsection{Feature Importance from Random Forests}


\begin{figure}
	\includegraphics[width=\columnwidth]{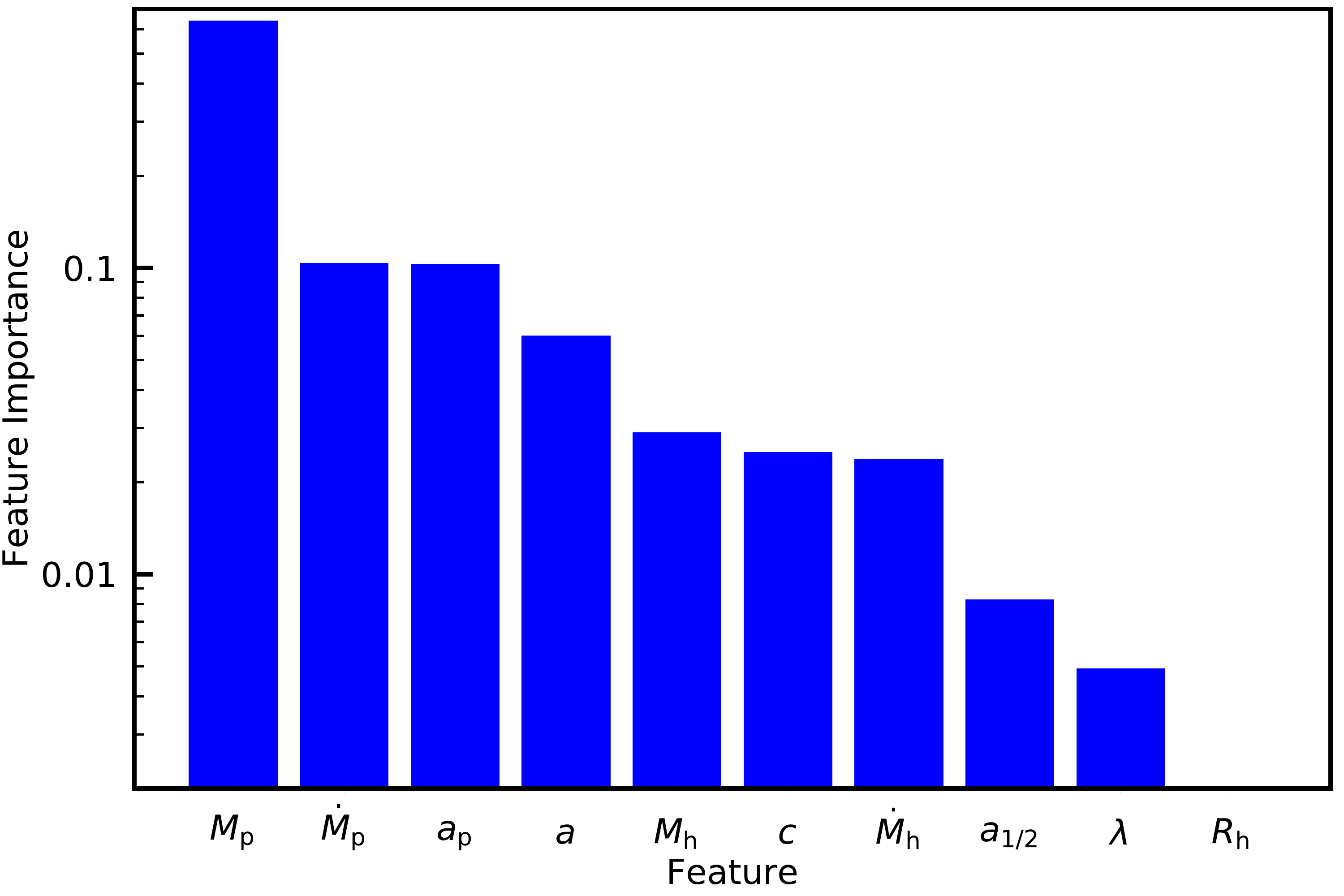}
    \caption{
        Feature importance as derived by a random forest. The height of each bar corresponds to the relative fraction of times a specific feature has been used in a tree to infer the targets (galaxy properties).
    }
    \label{fig:RF_feature_importance}
\end{figure}

With the scaled features and labels we can now perform a feature importance analysis. For this we use random forests, as this method includes a very simple and intuitive way to judge the relative importance of each feature. Random forests are based on decision trees, which can be grown with the classification and regression tree \citep[CART;][]{Breiman:1984aa} algorithm. For this, the training data is first split into two subsets using a single feature and a threshold, such that the two subsets have the minimum possible impurity, i.e. the difference between their mean label values and the individual labels in the sets. Once the data has been split in two subsets, the algorithm recursively splits each subset according to the same procedure. It stops recursing once the maximum depth (defined by a hyperparameter) has been reached, or if another stopping condition defined with a different hyperparameter, e.g. a minimum number of instances per subset. Once the tree has been grown, the prediction for each label is simply the mean label value of the deepest node (leaf) a training instance falls in. Care has to be taken to avoid overfitting, which can be prevented by limiting the depth or the minimum number of instances per leaf. As decision trees can be followed easily for each data instance, they provide a very intuitive way to judge which features have been used the most to decide in which leaf the instance ends up.


\begin{figure}
	\includegraphics[width=\columnwidth]{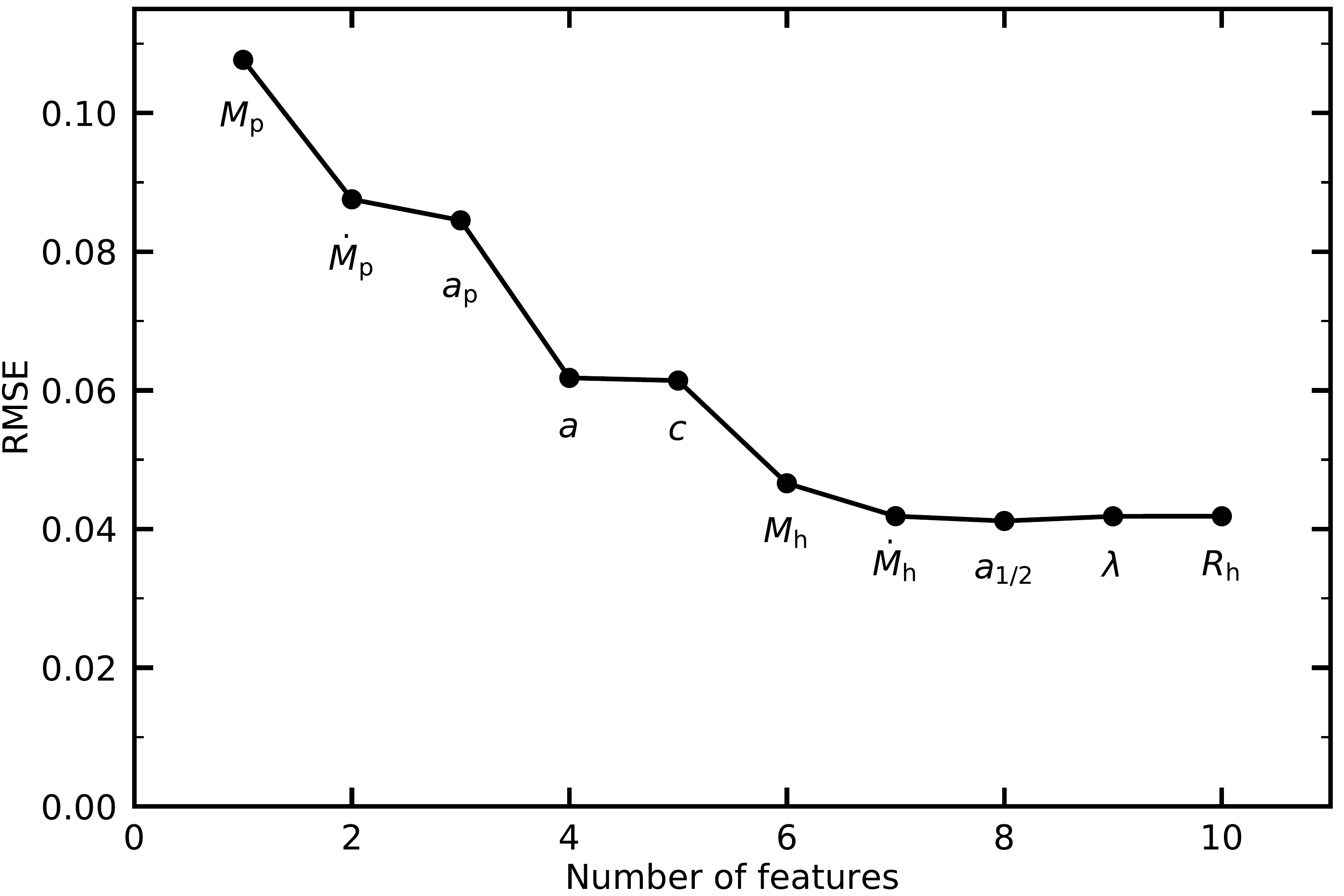}
    \caption{
        The root mean squared error as a function of the number of features used for the random forests. The labels for each point indicate the additional features used, i.e. the first feature is the halo peak mass, the second feature is the growth rate when the peak mass is reached, an so forth.
    }
    \label{fig:RF_rmse_vs_nfeatures}
\end{figure}


\begin{figure*}
	\includegraphics[width=\fullwidth]{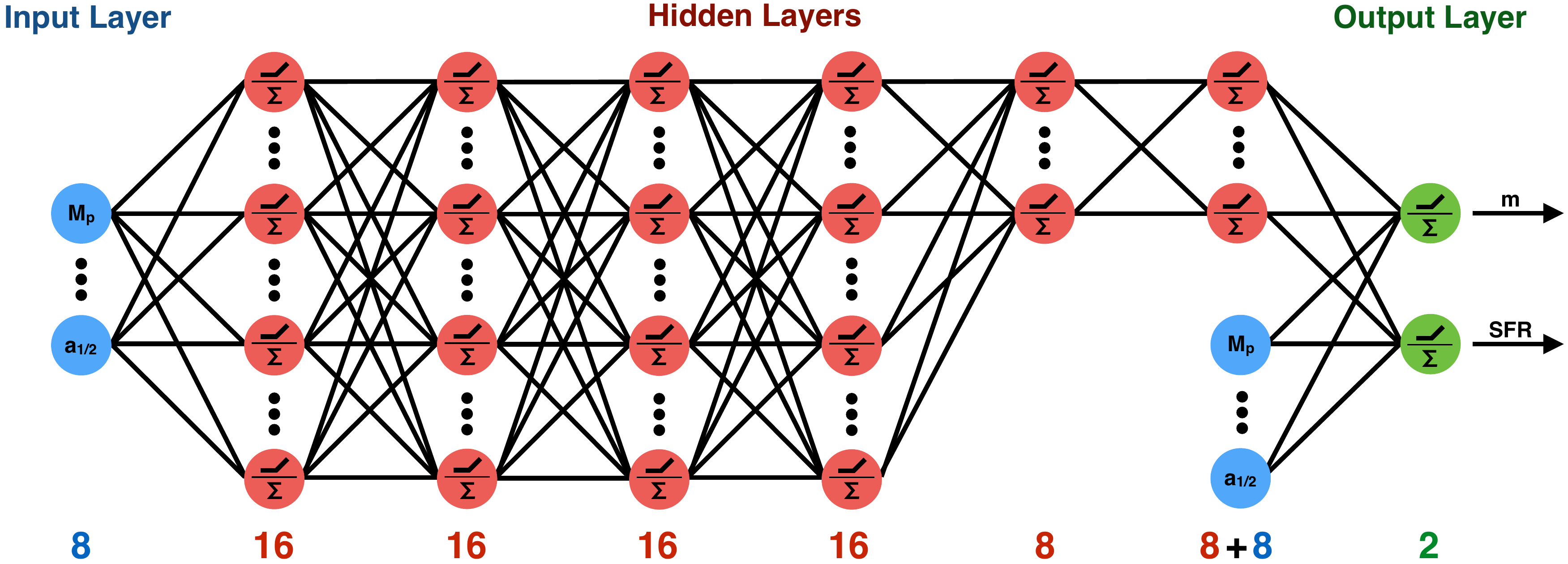}
    \caption{
        Architecture of the adopted wide \& deep neural network (WDNN). The input layer consists of 8 features, i.e. halo properties (blue). The first 4 hidden layers consist of 16 nodes, while the last two hidden layers have 8 nodes (red). The last hidden layer is concatenated with the 8 input features. The output layer gives the two targets, i.e. galaxy properties (green). All hidden layers are densely connected. The optimal number of nodes per layer is given at the bottom.
    }
    \label{fig:WDNN}
\end{figure*}

While decision trees are simple and intuitive, in practice methods perform better that aggregate the predictions of a group of regression models, called an ensemble. In ensemble learning, the final prediction is then given as the average of the prediction of each individual model. Random forests \citep{Breiman:2001aa} use decision trees as the individual regression models, and train each tree on a different random subset of the training data. This data sampling can either be done with replacement using bootstrap aggregation (bagging), or without replacement (pasting). Both techniques lead to a higher accuracy and a lower variance. Like decision trees, random forests allow for an easy measurement of the relative importance of each feature. This is achieved by combining the fraction of instances that are split according to a specific feature with the decrease in impurity that splitting with respect to this feature achieves. The estimates for the predictive ability of each feature is then averaged over several randomised trees, known as mean decrease in impurity \citep[MDI;][]{Louppe:2014aa}.

We use the \texttt{Scikit-learn} library \citep{Pedregosa:2011aa} to train the random forests for our data set, and to do the feature importance analysis. To determine the optimal hyperparameters for our problem, we first perform a grid search over the following four hyperparameters: the number of trees for the forest (\texttt{n\_estimators} = 16, 96, 512), number of features to consider when splitting (\texttt{max\_features} = 4, 8, \texttt{auto}), the minimum number of instances required to split a node (\texttt{min\_samples\_split} = 4, 6, 8), and whether to use bootstrapping or not (\texttt{bootstrap} = \texttt{True}, \texttt{False}). For the other hyperparameters, we use the default values of the \texttt{RandomForestRegressor} class. Based on the validation set, we find the best hyperparameters to be \texttt{n\_estimators} = 96, \texttt{max\_features} = \texttt{auto} (i.e. all features), \texttt{min\_samples\_split} = 6, and \texttt{bootstrap} = \texttt{False}.

Using these hyperparameters we train a random forest with all 10 features, and find the relative feature importance as shown in Figure \ref{fig:RF_feature_importance}. According to this analysis, the most important feature is the halo peak mass. This is not surprising as it is well known to be the most important halo property to predict the stellar mass of the galaxy in simple empirical models such as subhalo abundance matching. With a relative importance of 64 per cent, it is far more predictive than the current halo mass with only 3 per cent. The physical reason for this is simple: while the halo grows, the stellar mass of the galaxy is tightly correlated with the virial mass as it provides the potential well for gas to cool and thus determines the amount of gas that can fall to the centre and form stars. However, once the halo stops growing and starts to lose mass due to tidal stripping while orbiting in a larger host halo, the stellar mass of the galaxy does not decrease at the same rate as the halo mass. Instead it typically stays relatively constant, as the stripping of stars and star formation are of the same order. This implies that the current stellar mass is instead much more strongly correlated with the peak mass of the halo, as is recovered by the machine learning algorithm.

Similarly, we observe that the second and third most important features are the halo growth rate at the time the halo mass peaks (10 per cent), and the scale factor corresponding to this peak (10 per cent). This is also not particularly surprising, as the SFR in \textsc{emerge} depends on the growth rate of the halo, but after the halo mass peaks, it either stays constant, or is set to zero after the quenching time elapses. Consequently, the random forest needs the growth rate at the time the halo mass peaks to judge the SFR of a galaxy and the scale factor corresponding to the peak along with the present scale factor (6 per cent), to judge whether the quenching time has elapsed, and the SFR needs to be set to the minimum value. The current halo mass (3 per cent), the concentration (2.5 per cent), the current growth rate (2 per cent), and the half-mass scale factor (1 per cent) are of lesser importance, but are occasionally used to split the data sets. The halo spin is used even less with only 0.5 per cent relative importance, which signifies that it is irrelevant to predict stellar mass and SFR. However, we note that it could be far more important if the task were to predict galaxy sizes. Finally, we find that the virial radius is never used to split a data set and consequently has a relative importance of 0 per cent. This result is also expected, as the radius is monotonically related to the virial mass with $R_\mathrm{h} \propto \sqrt[3]{M_\mathrm{h}}$ without scatter.

As a further strategy to judge the most important features, we remove the least important feature (i.e. the virial radius), and repeat the feature importance analysis, i.e. we grow a random forest based on only 9 features and then calculate the relative importance of each. After this analysis, we again remove the least important feature, and repeat the procedure iteratively until only one feature is left. In Figure \ref{fig:RF_rmse_vs_nfeatures}, we show the resulting root mean squared error (RMSE) for each iteration given its number of features. The label below each point indicates the feature that is dropped in the next iteration. While not formally required, the resulting order of the features corresponds mostly to the one derived when all features were analysed at the same time. Only the concentration and the current halo mass switch their order, which is not surprising as they had almost the same relative importance. This confirms our assessment that the peak halo mass is the most important feature to determine the stellar mass and SFR of the galaxy, followed by the growth rate and scale factor at peak time, and the current scale factor. With these four features it is already possible to calculate the stellar mass and SFR to 6 per cent accuracy. Including additional features can decrease the RMSE further, but not below 4 per cent.


\section{Supervised Learning with a WDNN}
\label{sec:supervised}

Random forests are among the most powerful machine learning algorithms and generally show a performance comparable or superior to other methods \citep{Fernandez:2014aa,Ahmad:2017aa,Nawar:2017aa}. However, training the model with reinforcement learning is very complex with random forests, as the parameter space does not have a fixed dimensionality. Consequently, optimisation is not straight forward and can be very slow. Therefore, we choose neural networks to learn the connection between halo and galaxy properties. They have a fixed number of parameters, so that the model can be optimised very efficiently with stochastic methods. Having established an importance ranking for the features, we can now proceed to determine the optimal network architecture and hyperparameters. As we need to evaluate the performance with a validation set, we investigate different setups with a standard supervised learning approach. Once the optimal network has been determined to predict the galaxy properties, we will employ this network in a reinforcement learning approach.

In section \ref{sec:featureselection}, we found that the target values can be predicted very accurately with only 6 features, which results in an RMSE that is almost the same as if one uses all 10 features (Figure \ref{fig:RF_rmse_vs_nfeatures}). However, we carry out all computations on NVIDIA tensor cores, which are optimised for numbers of input features that are multiples of 8. Consequently, we use the 8 most important features to train the neural networks, i.e. all features except the spin and the radius. All results and figures in this work are based on this choice. However, we also evaluate the network results based on only the four most important features (peak mass, peak growth rate, peak scale factor, and current scale factor), and report our findings as a reference.


\subsection{GalaxyNet: A Wide \& Deep Neural Network}


\begin{table}
	\centering
	\caption{Architecture of the adopted wide \& deep neural network (WDNN). The columns are (1) layer name, (2) previous layer connected to, (3) number of nodes in this layer, (4) activation function, and (5) number of trainable parameters.}
	\label{tab:architecture}
	\begin{tabular}{llccr}
		\hline
		Layer & Input & Nodes & AF & Parameters\\
		\hline
		Input & -- & 8 & -- & 0\\
		Hidden-1 & Input & 16 & Selu & 144\\
		Hidden-2 & Hidden-1 & 16 & Selu & 272\\
		Hidden-3 & Hidden-2 & 16 & Selu & 272\\
		Hidden-4 & Hidden-3 & 16 & Selu & 272\\
		Hidden-5 & Hidden-4 & 8 & Selu & 136\\
		Hidden-6 & Hidden-5 & 8 & Selu & 72\\
		Concat & Input \& Hidden-6 & 16 & -- & 0\\
		Output & Concat & 2 & Selu & 34\\		
		\hline
		\multicolumn{4}{l}{Total trainable parameters:} & 1,202\\
		\hline
	\end{tabular}
\end{table}

Artificial neural networks exist in many flavours, such as convolutional networks or recurrent networks. Here we employ one of the most simple architectures, the multi-layer perceptron (MLP). This type of network consists of an input layer with one node per feature, several hidden layers with a varying number of nodes, and an output layer with one node per target. All nodes in a layer are connected to every node in the previous layer, such that all layers are \textit{fully connected} or \textit{dense} layers. Each node then performs two calculations: first it computes a weighted sum of its inputs, and second it applies an \textit{activation function} $\phi$ to that sum, which leads to non-linearity in the model. For a whole layer, this can be written as a simple matrix multiplication:
\begin{equation}
\mathbfit{y} = \phi \, (\mathbfss{W} \, \mathbfit{x} + \mathbfit{b}) \; .
\end{equation}
Here, \mathbfit{y} represents the vector of outputs of this layer, $\mathbfit{x}$ represents the vector of inputs, i.e. the outputs of the previous layer, $\mathbfss{W}$ is the weight matrix containing all connection weights $w_{ij}$, and $\mathbfit{b}$ is the bias vector containing one bias term for each node.

While this architecture already leads to good results, a regular MLP forces all the data to flow through the full stack of layers. The disadvantage of this approach is that simple patterns in the data may end up being distorted by this sequence of transformations. Therefore we use a variation of a standard MLP as the general structure for GalaxyNet: a wide \& deep neural network \citep[WDNN;][]{Cheng:2016aa}. It contains a deep part, which is achieved with a standard MLP, and a wide part, which connects the features directly to the output layer. This architecture allows the neural network to learn both deep patterns (through the deep path), and simple rules (with the short path). A typical WDNN architecture is shown in Figure \ref{fig:WDNN}, where the input layer containing the features is given by the blue nodes, the hidden layers are given by the red nodes and the output layer containing the targets is given by the green nodes. All hidden layers are fully connected, and the last hidden layer is concatenated with the input layer. Each node then first performs a weighted sum, and then applies the activation function.


\begin{figure}
	\includegraphics[width=\columnwidth]{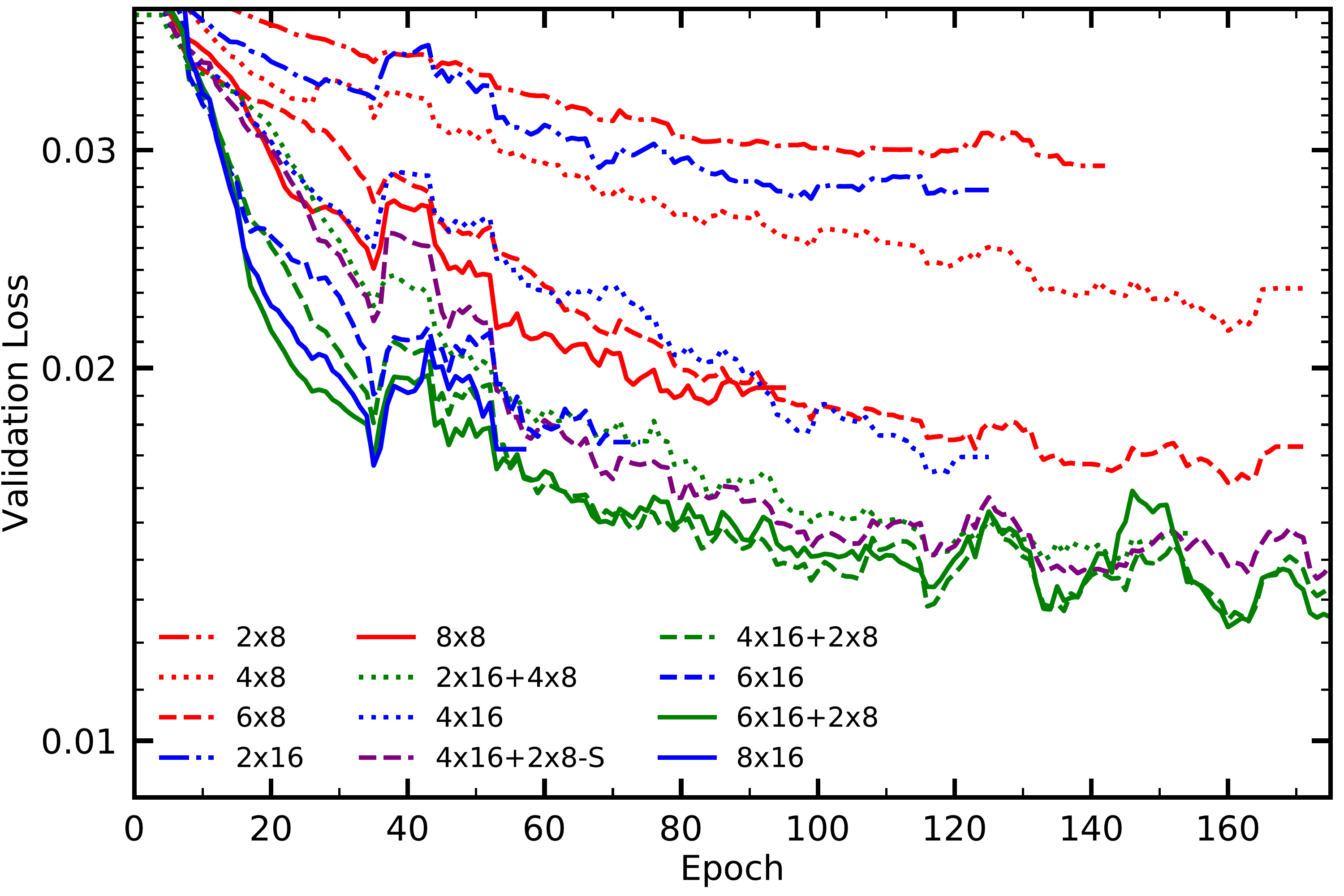}
    \caption{
        Training histories for all tested neural network architectures. All models are wide \& deep networks, except for one standard deep network (4x16+2x8-S).
    }
    \label{fig:DNN_training_history}
\end{figure}

For the activation function, we use scaled exponential linear units \citep[SELUs;][]{Klambauer:2017aa} which take the form
\begin{equation}
\phi (x) = \lambda \begin{cases} x &\mbox{if } x > 0 \\ 
\alpha e^x - \alpha & \mbox{if } x \le 0 \end{cases} \; ,
\end{equation}
where $\lambda=1.05$ and $\alpha=1.67$ are chosen so that the mean and variance of the inputs are preserved between two consecutive layers. In this way, SELUs lead to self-normalising neural networks, such that the output of each layer will tend to preserve a mean of 0 and standard deviation of 1 during training, which solves the vanishing and exploding gradients problem. Moreover, unlike rectified linear units (ReLUs) they do not result in dying nodes, and they tend to lead to faster learning. To facilitate the training with SELUs we use a \citet{LeCun:1998aa} normal initialisation for the weight matrices. Although the features have not been standardised and our WDNN has a non-sequential architecture with skip connections, we found the SELU activation function to perform better than others like ReLU, leaky ReLU, and ELU, with a faster convergence. We further tested batch normalisation and regularisation (L1 and L2), but found a better convergence and performance on our validation set without applying these techniques. The code has been implemented in python using the TensorFlow library in version 2.1 \citep{Abadi:2015aa,Abadi:2016aa} with the Keras API \citep{Chollet:2015aa}.


\subsection{Training GalaxyNet with backpropagation}

All networks are trained on two NVIDIA Volta (V100) GPUs employing the tensor cores with mixed precision training. For the supervised learning approach in this section, which we employ to find the optimal network architecture and hyperparameters, we use the standard backpropagation algorithm \citep{Rumelhart:1986aa}. The target labels are provided by \textsc{emerge} as described in section \ref{sec:labels}. For the loss function we use the mean absolute error (MAE), which showed a better performance than the mean squared error (MSE). One complication with using supervised learning for galaxies is their mass distribution, i.e. there are far more low-mass galaxies than massive ones. Consequently, the training will concentrate on the low-mass galaxies and try to get the lowest possible error for them. The error for massive galaxies may then be much larger, as for the total error they would be negligible. One solution to this problem would be to use stratified sampling, but this reduces the size of the training set significantly. Therefore we opted for an alternative and weight the error of each data instance by how often a galaxy with its stellar mass appears. In practice this means that we multiply each individual error with a weight that is proportional to the inverse of the stellar mass function for the stellar mass that corresponds to the label, with a mean weight of 1. To train the parameters, we use the \textit{adaptive moment estimation} \citep[Adam;][]{Kingma:2014aa} optimiser with a learning rate of 0.001. We use a batch size of $10,000$ and apply early stopping with a patience of 20, i.e. we stop training once the validation error has not decreased in the last 20 epochs. 

With the fixed input and output layer sizes, we tested various architectures for the hidden layers: 2, 4, 6, and 8 layers with either 8 or 16 nodes each (2x8, 4x8, 6x8, 8x8, 2x16, 4x16, 6x16, 8x16), as well as mixed setups that have 2 layers with 16 nodes followed by 4 layers with 8 nodes (2x16+4x8), or 4 layers with 16 nodes followed by 2 layers with 8 nodes (4x16+2x8), and finally 6 layers with 16 nodes followed by 2 layers with 8 nodes (6x16+2x8). All networks use the wide \& deep architecture, i.e. the features are concatenated with the last hidden layer, except for one additional network, that only uses a standard MLP architecture (4x16+2x8-S). Figure \ref{fig:DNN_training_history} shows the training histories, i.e. the validation loss as function of the epoch, for each network. We find that the more complex networks with more free parameters tend to overfit relatively early-on, especially the ones that have the same number of nodes for each layer. On the other hand, simple networks with only few hidden layers perform poorly as well. Networks with a reversed pyramid shape topology perform much better for our data set.


\begin{figure}
	\includegraphics[width=\columnwidth]{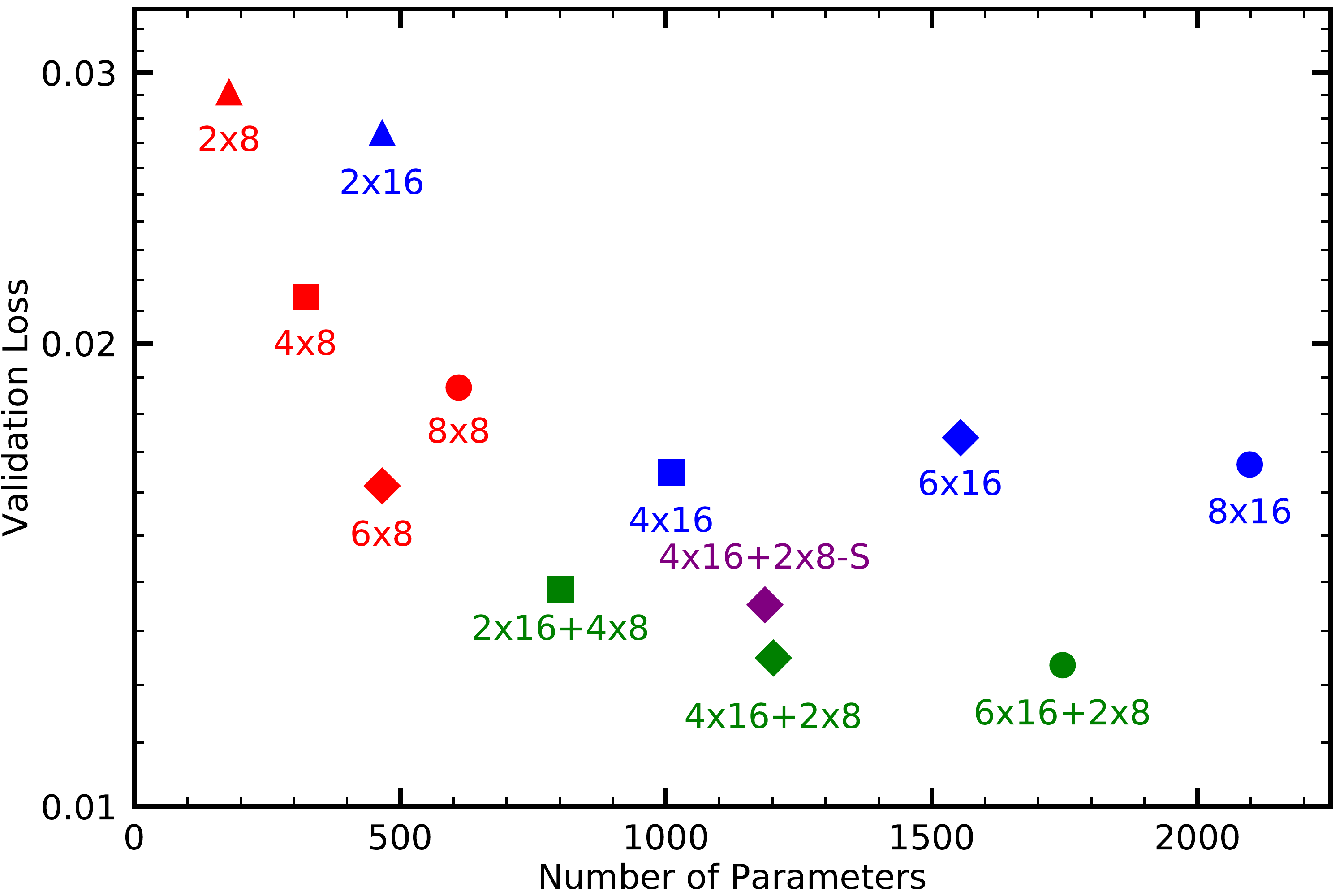}
    \caption{
        Best validation loss as a function of the number of model parameters. Networks with too little complexity (few layers and nodes) tend to underfit, while too complex models can start to overfit early-on. The best performance is achieved with networks that have a reversed pyramid shape topology (4x16+2x8 and 6x16+2x8).
    }
    \label{fig:DNN_loss_vs_nparam}
\end{figure}

In Figure \ref{fig:DNN_loss_vs_nparam}, we show the best validation loss achieved with each network as a function of the number of model parameters (weights and biases). The red, blue, and green symbols are for networks with 8, 16, and a mixed number of nodes per layer. The triangles, squares, diamonds, and circles are for networks with 2, 4, 6, and 8 hidden layers. The networks with the lowest loss on the validation set have an architecture with 4 or 6 layers of 16 nodes and 2 final layers with 8 nodes. Their best validation loss values are at a comparable level with 1.25 per cent for the 4x16+2x8 network and 1.24 per cent for the 6x16+2x8 network, using the MAE for the loss. Since the number of trainable parameters is considerably lower for the 4x16+2x8 network with 1,202 parameters, compared to the 6x16+2x8 network with 1,746 parameters, we select the 4x16+2x8 network as our default architecture for GalaxyNet. We find similar results for the smaller network with only 4 features, and determine the best network to have 4 layers with 8 nodes followed by 2 layers with 4 nodes (4x8+2x4), with a validation loss of 1.39 per cent.


\begin{figure}
	\includegraphics[width=\columnwidth]{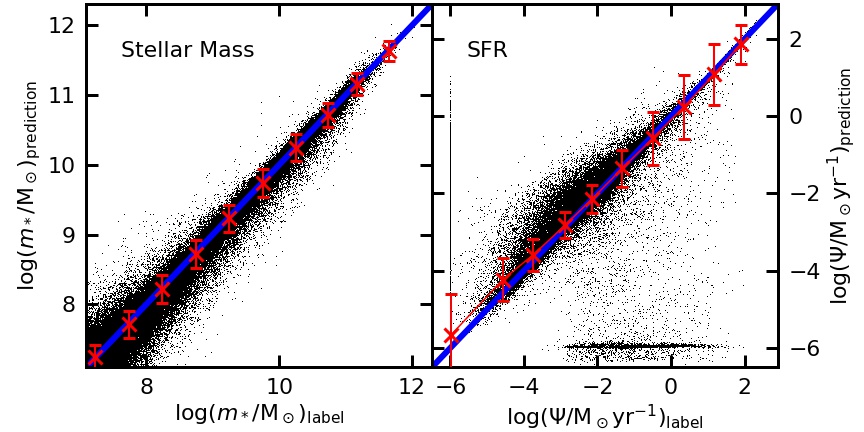}
    \caption{
        Comparison between target labels and network prediction for all galaxies. The left-hand panel shows the stellar mass, while the right-hand panel show the SFR. The blue line indicates the perfect prediction (one-to-one), while the red symbols show the average and standard deviation of the prediction for fixed label bins. 
    }
    \label{fig:GalaxyNet_lab_vs_pred}
\end{figure}


\subsection{Resulting Galaxy Properties}

Having found the optimal network architecture for our problem, we can now investigate how well the network is able to reproduce the galaxy properties. We compare the model predictions with the labels in Figure \ref{fig:GalaxyNet_lab_vs_pred}. The left-hand and right-hand panels show the results for the stellar mass and SFR, respectively. Each point corresponds to one galaxy, with the horizontal axis showing the label from \textsc{emerge} (the `true' value), and the vertical axis showing the network prediction. The blue line indicates a perfect prediction, i.e. a one-to-one agreement. The red symbols give the mean value for the prediction in fixed label bins, and the error bars give the standard deviation. For the stellar mass we find a very good agreement with the labels independent of mass or redshift. The deviation of the mean prediction from the label is typically smaller than 5 per cent, and the standard deviation is of the order of 0.2 dex.

Reproducing the SFR with high accuracy proves to be more difficult. For most galaxies this works very well, with a typical deviation from the labels of 20 per cent, and a standard deviation of 0.3 dex. However, we find that the neural network has trouble modelling galaxies that are completely quenched and have the minimum SFR. There several galaxies that either have a label with the minimum value but a prediction that is much higher (points on the left of the right-hand panel of Figure \ref{fig:GalaxyNet_lab_vs_pred}), or have a label with a relatively high label value but a prediction that has the minimum SFR (points on the bottom of the right-hand panel of Figure \ref{fig:GalaxyNet_lab_vs_pred}). The difficulty lies in the method \textsc{emerge} uses to decide when satellite galaxies become quenched (delayed-then-rapid). In \textsc{emerge}, a galaxy in a halo that has started to lose mass keeps forming stars at its current rate until the quenching time has elapsed, after which the SFR is set to zero.

To reproduce this, the neural network has to compare the scale factor at peak mass $a_\mathrm{p}$ to the current scale factor $a$ and decide whether to use the growth rate at peak mass $(\mathrm{d} M / \mathrm{dt})_\mathrm{p}$ to compute the SFR, or whether to set the SFR to the minimum value. As the quenching time depends on redshift and stellar mass in \textsc{emerge}, $a-a_\mathrm{p}$ is sometimes not enough to determine if a galaxy has been quenched and it will be misclassified resulting in the wrong SFR. If the galaxy in \textsc{emerge} is still forming stars but is about to be quenched soon, and the network decided to already quench it, it ends up on the bottom of plot. On the other hand, if the galaxy had just been quenched in \textsc{emerge}, but the network has not decided to quench it just yet, it ends up on the left of the plot. Still, these misclassified galaxies are only a small fraction, such that the average of the prediction is very close to the label value. 


\begin{figure*}
	\includegraphics[width=\fullwidth]{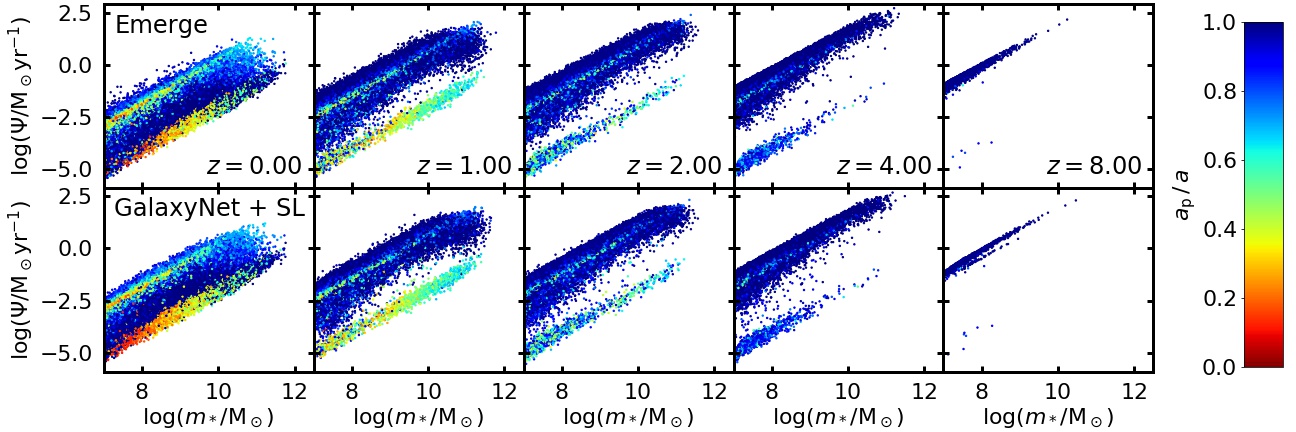}
    \caption{
        Comparison of the SFR vs stellar mass relation between \textsc{emerge} (top panels) and GalaxyNet trained with supervised learning (bottom panels). From left to right the panels show the results at five different redshifts. The colour of each point corresponds to the ratio between the scale factor at peak mass and the scale factor at this redshift, i.e. for blue points the halo mass peaks at the current redshift, while for red points the halo mass peaked early-on. For both models the observed main sequence and quenched cloud are well reproduced. The galaxies on the main sequence are predominantly in haloes that are currently at their peak mass, while the quenched galaxies typically live in haloes that peaked in mass at early times. 
    }
    \label{fig:GalaxyNet_SL_MS}
\end{figure*}


\begin{figure*}
	\includegraphics[width=\fullwidth]{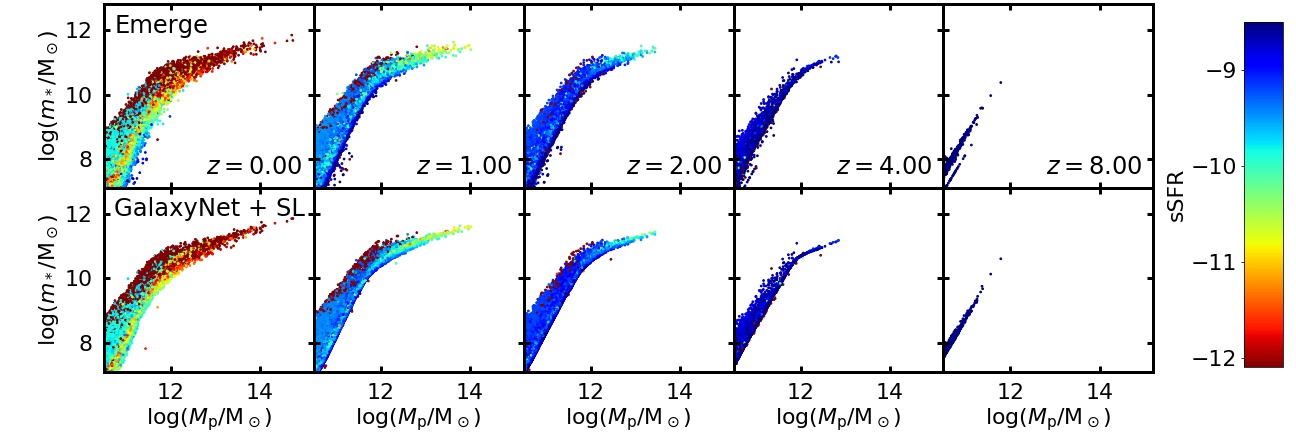}
    \caption{
        Comparison of the stellar-to-halo mass (SHM) relation between \textsc{emerge} (top panels) and GalaxyNet trained with supervised learning (bottom panels). From left to right the panels show the results at five different redshifts. The colour of each galaxy corresponds its specific SFR, i.e. blue galaxies are actively forming stars, while red galaxies are passive. The predictions of GalaxyNet agree very well with \textsc{emerge} and demonstrate that galaxies in massive haloes tend to be passive, and that at fixed halo mass, passive galaxies tend to have more stellar mass than active galaxies.
    }
    \label{fig:GalaxyNet_SL_SHMR}
\end{figure*}

With the neural network predictions being in good agreement with the labels, we can now compare \textsc{emerge} with GalaxyNet trained with supervised learning in more detail. Figure \ref{fig:GalaxyNet_SL_MS} compares the two targets, i.e. the relation between the SFR and the stellar mass for each galaxy. The top panels show the labels by \textsc{emerge}, while the bottom panels show the predictions by GalaxyNet, each for five different redshifts. The colour of each point indicates the ratio $a_\mathrm{p}/a$ between the scale factor at peak mass $a_\mathrm{p}$ and the scale factor at this redshift $a$. For blue points the halo mass reaches its peak at the current redshift (typical for central galaxies), while for red points the halo mass peaked much earlier (typical for satellites). Both \textsc{emerge} and GalaxyNet are able to reproduce the observed main sequence and the quenched cloud very well. We find that main sequence galaxies are predominantly in haloes that are currently at their peak mass, although at low redshift there are a few galaxies on the main sequence with a halo that reached its peak mass much earlier. The quenched cloud mainly consists of satellite galaxies for which the halo mass peaked much earlier. However, at the massive end of the quenched cloud, the majority of galaxies are centrals with halo masses that currently peak. This indicates that low-mass galaxies are mainly quenched because of becoming satellites and hence losing their gas, while massive galaxies tend to get quenched because of internal processes (feedback).

A second comparison can be made between the two models for the relation between peak halo mass and stellar mass (the SHM relation), which we show in Figure \ref{fig:GalaxyNet_SL_SHMR}. As above, each point represents a galaxy with the colour corresponding to its specific SFR (sSFR). Blue galaxies with high sSFRs are actively forming stars, while red galaxies with low sSFRs are passive. Both models agree very well and show that galaxies in massive haloes tend to be passive. Moreover, at fixed halo mass, passive galaxies typically have a higher stellar mass than active galaxies. Although the predictions by GalaxyNet show a little less scatter, we find that at fixed stellar mass, passive galaxies tend to be hosted by more massive haloes compared to blue galaxies, in agreement with lensing predictions (see \citealt{Moster:2019aa} for more details on this).


\section{Reinforcement Learning}
\label{sec:reinforcement}

In the previous section we showed that it is possible to train a wide \& deep neural network to reproduce the relation between halo and galaxy properties with supervised learning. However, the best possible result is to reproduce the data that the network has been trained on, which comes from some other model. This is problematic for two reasons. First, whatever model has been used to train the network comes with its own limitations and problems, so it is not desirable to reproduce these. Second, the model has already been established and fitted, so that simply reproducing it is often not useful and does not provide any advantage (one should be very careful with extrapolating any machine learning model, e.g. beyond the trained masses). The main goal of this paper is therefore not to emulate any other model, but to directly train a neural network with observed data. Obtaining reliable labeled data (with galaxy and halo properties for each instance) directly from observations is not possible with the current means. Consequently, we train GalaxyNet on a number of observed statistical data sets, such as the SMF, local clustering, cosmic and specific SFRs, and quenched fractions.

One of the most powerful strategies we can apply when no labelled data is available is reinforcement learning. Classically, in reinforcement learning an \textit{agent} takes actions within some environment. Based on its actions, it receives rewards and must learn to act in a way that will maximise its rewards over time, i.e. it must optimise a global problem. The algorithm that the agent uses to determine its actions is called the \textit{policy}. Unlike in supervised learning, the agent is not explicitly given the `correct' answer (label), but must learn by trial and error. In contrast to unsupervised learning though, there is a form of supervision, through rewards. The agent is not told which actions to perform, but when it is making progress and when not. Further, when exploring the environment or parameter space, a reinforcement agent needs to find the right balance between finding new ways to gain rewards, and profiting from existing sources of rewards.

For our problem we can adapt a reinforcement learning strategy as follows. The halo properties define the environment, and the network GalaxyNet is the agent that takes actions in the form of deriving galaxy properties and predicting global statistics (such as SMFs). The results are then compared to the observed statistics with a loss function, which sets the reward, that can be maximised by minimising the loss. Finally, the policy is set by a stochastic optimisation algorithm, such as particle swarm optimisation, which tries to minimise the loss. This means that the PSO algorithm tries to find the optimal position in parameter space (spanned by network weights and biases) to gain the most reward, bringing the network prediction in the best possible agreement with the observations.


\subsection{Combining Different Observed Data Sets}

The first task to enable reinforcement learning is to obtain several observed data sets that can be used to compute the loss. To this end, we employ the same data that was used in \citet{Moster:2018aa} and updated in \citet{Moster:2019aa}. These include SMFs up to $z\sim8$, the fraction of quenched galaxies as function of stellar mass (FQ) up to $z\sim3$, sSFRs as function of stellar mass up to $z\sim8$, the cosmic SFR density as function of redshift (CSFRD) up to $z\sim11$, and local clustering for several stellar mass bins (WP). To be able to compare the neural network with these observations efficiently, we compute the average observed statistics over different data sets. For the SMFs, FQs, and the sSFRs, we bin the observations in stellar mass from $\log(m_*/\Msun) = 7.0$ to $12.4$ with $\Delta\log m_* = 0.4$, and in redshift centred around our 10 dark matter simulation outputs from $z=0$ to $8$. Similarly, for the CSFRD, we use the same redshift bins to compute an average. Finally, for WP we average the observed projected correlation functions for all four stellar mass bins in radial bins from 10 kpc to 50 Mpc with $\Delta\log r_\mathrm{p} = 0.01$. This procedure results in 520 averaged data points for the observed statistics.


\subsection{The Loss Function}

After each training epoch, i.e. once the network has computed the properties of all galaxies given a set of weights and biases, we use the resulting mock galaxies to compute model statistics. In every redshift - stellar mass bin, we count the number of galaxies (active and passive) to derive the SMFs and FQs, and we compute the average SFR to derive the sSFRs. Active and passive galaxies are separated by a sSFR threshold of $0.5t_\mathrm{H}^{-1}(z)$, where $t_\mathrm{H}(z)$ is the Hubble time at redshift $z$. We compute the sum of the SFR of all galaxies in each redshift bin to get the CSFRD. To compute the projected correlation functions in four stellar mass bins we use the Halotools v0.6 python library \citep{Hearin:2017aa}.

Once we have computed all statistics for the model galaxies, we can compute the loss $L$ which in turn sets the reward for the agent. The policy then explores the parameter space spanned by the network weights and biases $\bm{\theta}$, and tries to gain the largest possible reward by minimising the loss $L(\bm{\theta})$. We compare the observed data points $\bm{\omega}$ to the model values $\bm{\mu}(\bm{\theta})$ given the current parameters $\bm{\theta}$ and the uncertainty of the observed data $\bm{\sigma}$ to compute the loss
\begin{equation}
L(\bm{\theta}) = \sum_i \left( \frac{\omega_i - \mu_i(\bm{\theta})}{\sigma_i} \right)^2 \; .
\end{equation}


\subsection{Training with Particle Swarm Optimisation}


\begin{figure*}
	\includegraphics[width=\fullwidth]{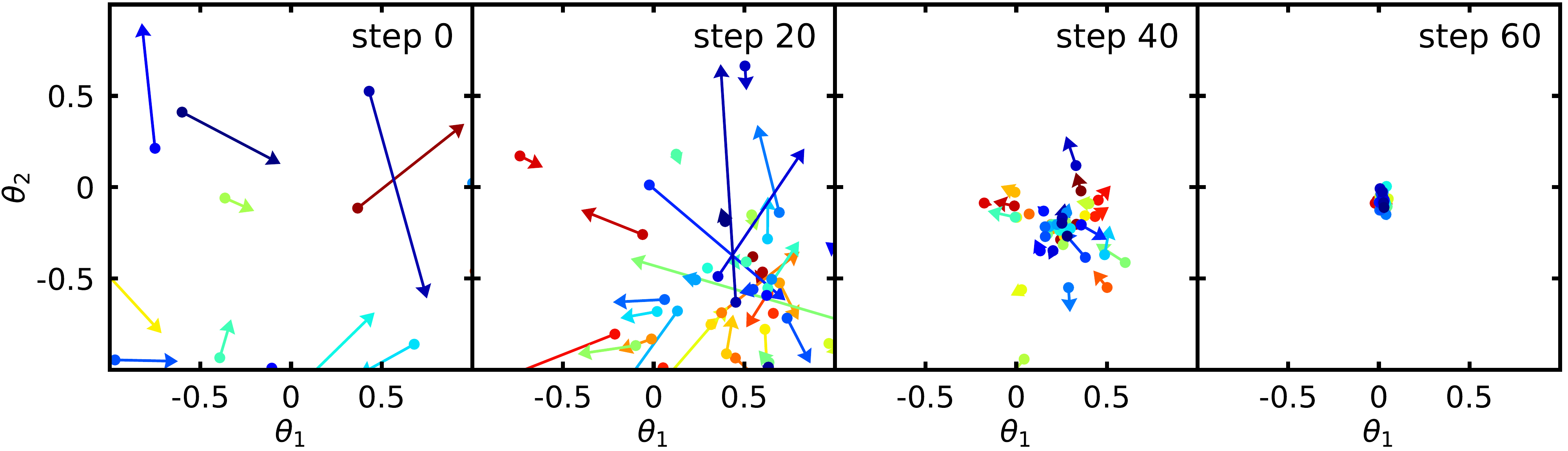}
    \caption{
        Illustration of the particle swarm optimisation algorithm for the 2 dimensions with the least variance. The parameters have been scaled such that the minimum is at the origin. The PSO step number is given in the top right corner of each panel. Each point corresponds to one particle with the arrow indicating the current velocity. While initially the particles explore a vast region of the parameter space, they finally converge to the global minimum.
    }
    \label{fig:pso}
\end{figure*}


\begin{table}
	\centering
	\caption{Description of the particle swarm optimisation algorithm. The hyperparameters have been set to $c_1=0.9$, $c_2=0.9$, $w_\mathrm{max}=0.9$, $w_\mathrm{min}=0.5$, $f_\mathrm{vel}=0.5$, and $\Delta L_\mathrm{stop}=0.5$.}
	\label{tab:pso}
	\begin{tabular}{lll}
		\hline
		\multicolumn{3}{l}{\textbf{Algorithm} -- Particle Swarm Optimisation}\\
		\hline
		& \multicolumn{2}{l}{Initialise $n$ particle positions $\mathbfit{x}_i^0$ for $i = 1$ to $n$}\\[2pt]
		& \multicolumn{2}{l}{Initialise velocities $\mathbfit{v}_i^0 = f_\mathrm{vel}\,(\mathbfit{x}_i^0-\mathbfit{x}_j^0)$ for random $j\neq i$}\\[2pt]
		& \multicolumn{2}{l}{Initialise particle best $L_{i,\mathrm{pb}} = \infty$ and swarm best $L_{\mathrm{sb}} = \infty$}\\[2pt]
		& \multicolumn{2}{l}{\textbf{for} $s$ in $N_\mathrm{steps}$ \textbf{do}:}\\[2pt]
		& & Set $w = w_\mathrm{max} - (w_\mathrm{max} - w_\mathrm{min})  \cdot (s / N_\mathrm{steps})$\\[2pt]
		& & Evaluate $L_i^s = L(\mathbfit{x}_i^s)$ for all $\mathbfit{x}_i^s$\\[2pt]
		& & \textbf{if} $L_i^s < L_{i,\mathrm{pb}}$ \textbf{then} $L_{i,\mathrm{pb}} = L_i^s$ and $\mathbfit{x}_{i,\mathrm{pb}} = \mathbfit{x}_i^s$\\[2pt]
		& & \textbf{if} $L_i^s < L_{\mathrm{sb}}$ \textbf{then} $L_{\mathrm{sb}} = L_i^s$ and $\mathbfit{x}_{\mathrm{sb}} = \mathbfit{x}_i^s$\\[2pt]
		& & Draw uniform random numbers $q_i$ and $r_i$ in $[0,1)$ for all $i$\\[2pt]
		& & Set $\mathbfit{v}_i^s = w \mathbfit{v}_i^{s-1} + c_1 q_i(\mathbfit{x}_{i,\mathrm{pb}} - \mathbfit{x}_i^{s-1}) + c_2 r_i (\mathbfit{x}_{\mathrm{sb}} - \mathbfit{x}_i^{s-1})$\\[2pt]
		& & Set $\mathbfit{x}_i^s = \mathbfit{x}_i^{s-1} + \mathbfit{v}_i^s$\\[2pt]
		& & Stop \textbf{if} $|\langle L_i^s \rangle - \langle L_i^{s-1} \rangle| < \Delta L_\mathrm{stop}$\\[2pt]
		& \multicolumn{2}{l}{Best position given by $\mathbfit{x}_{\mathrm{sb}}$}\\[2pt]
		& \multicolumn{2}{l}{Best loss given by $L_{\mathrm{sb}}$}\\[2pt]
		\hline
	\end{tabular}
\end{table}

The final aspect of the reinforcement learning approach is to set the policy that determines the actions of the agent in a way that maximises the reward and minimises the loss. To this end, we employ particle swarm optimisation \citep{Eberhart:1995aa}, which is a stochastic method based on an ensemble of particles called a swarm. Each particle is located in an $N$-dimensional parameter space with the position given by the parameter values. Further, each particle has a velocity, so that it can move through parameter space and compute the loss. Initially, the swarm particles are distributed over a large region, but with time they close in on the areas with the lowest loss to eventually find the global minimum. To facilitate this, each particle memorises the best loss that has been obtained so far, both by the particle itself and by the swarm as a whole. With this information, the velocity of each particle is updated each step, such that they swarm around these two best locations.

The details of the PSO algorithm are described in Table \ref{tab:pso}. We create a swarm with $n=50$ particles, and initialise the positions in a large region around a specific location. The initial velocities are set, such that each particle moves half of the distance towards some other random particle during the first step ($f_\mathrm{vel}=0.5$). The inertia weight $w$ ensures that the particles overshoot the position they are moving towards. We initialise it with $w_\mathrm{max}=0.9$ and decrease it to $w_\mathrm{min}=0.5$, which initially favours exploration of unknown regions of parameter space and later on exploitation of space now known to return a high reward. The parameters $c_1$ and $c_2$ regulate the amount of focus placed on gravitating toward the particle best or swarm best, called the \textit{cognitive} and \textit{social} components, respectively. The sum of $c_1$ and $c_2$ is constrained by $c_1 + c_2 \le 4$, to prevent the trajectories of the particles from becoming unbound. We found best convergence for $c_1 = c_2 = 0.9$. The random numbers $q$ and $r$ are drawn at every step for each particle from the uniform distribution $[0, 1)$. We stop the PSO algorithm once the mean loss of the swarm does no longer change by more than the value $\Delta L_\mathrm{stop}=0.5$.

An illustration of the PSO algorithms for the two dimensions with the least variance is shown in \ref{fig:pso}, and demonstrates that the optimiser finds the minimum after $\sim60$ steps, which corresponds to $\sim3000$ calls to the loss function. We find a final loss value of $L_\mathrm{PSO} = 544$, which is significantly lower than the loss we compute for \textsc{emerge} with $L_\mathrm{EM} = 2220$, and the loss for the supervised learning trained GalaxyNet with $L_\mathrm{SL} = 2268$. For the network with only 4 features, we still find a good final loss of $L_\mathrm{PSO,4F} = 663$. This result shows that training a neural network with reinforcement learning can reproduce the observed statistics very well. We now investigate how the statistics have been improved with respect to the empirical model.


\subsection{Resulting Galaxy Properties and Statistics}


\begin{figure*}
	\includegraphics[width=\fullwidth]{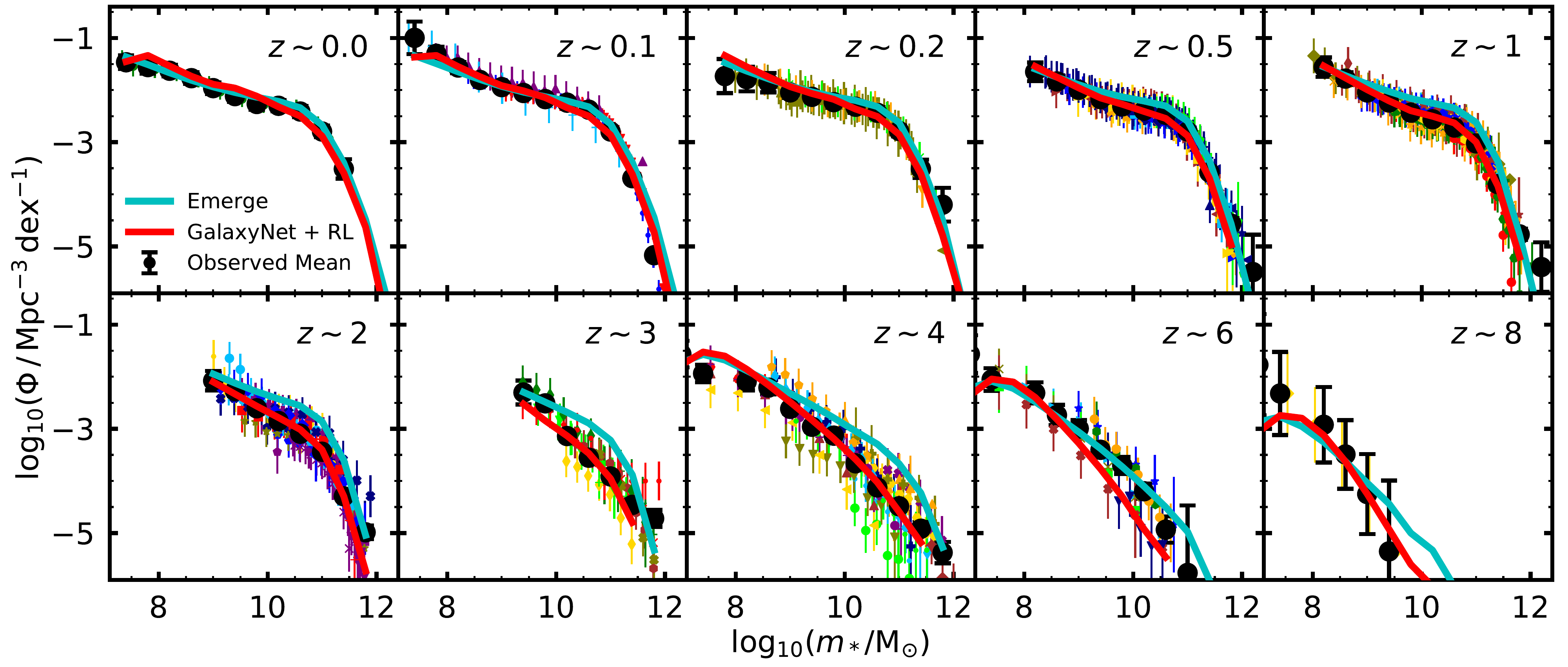}
    \caption{
        Comparison of the stellar mass function between \textsc{emerge} (blue lines) and GalaxyNet trained with reinforcement learning (red lines) at ten redshifts. The coloured symbols in the background correspond to the original observed data sets use to create the mean observed values (black symbols with error bars).
    }
    \label{fig:GalaxyNet_RL_SMF}
\end{figure*}


\begin{figure*}
	\includegraphics[width=\fullwidth]{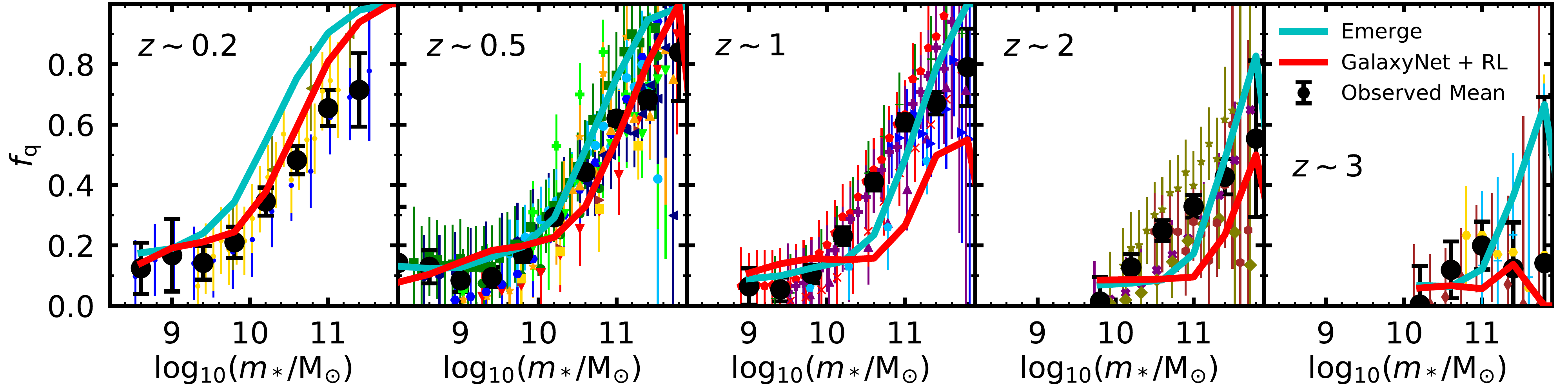}
    \caption{
        Comparison of the fraction of quenched galaxies as function of stellar mass between \textsc{emerge} (blue lines) and GalaxyNet trained with reinforcement learning (red lines) at five redshifts. The coloured symbols in the background correspond to the original observed data sets use to create the mean observed values (black symbols with error bars).
    }
    \label{fig:GalaxyNet_RL_FQ}
\end{figure*}


\begin{figure*}
	\includegraphics[width=\fullwidth]{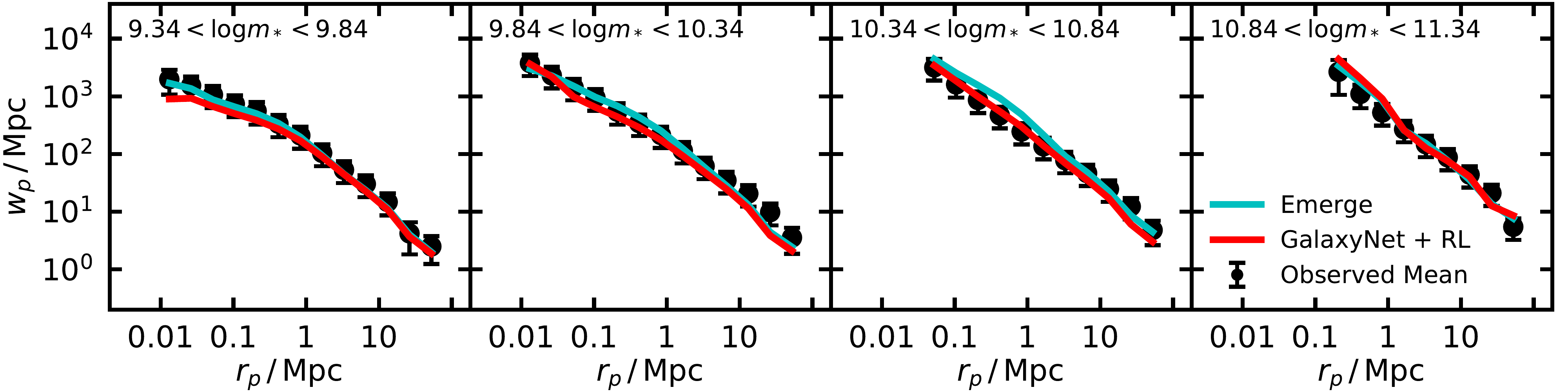}
    \caption{
        Comparison of the projected two-point correlation function between \textsc{emerge} (blue lines) and GalaxyNet trained with reinforcement learning (red lines) for four stellar mass bins. The black symbols with error bars show the mean observed values. The correlation functions have been computed at $z=0.1$.
    }
    \label{fig:GalaxyNet_RL_WP}
\end{figure*}


\begin{figure*}
	\includegraphics[width=\fullwidth]{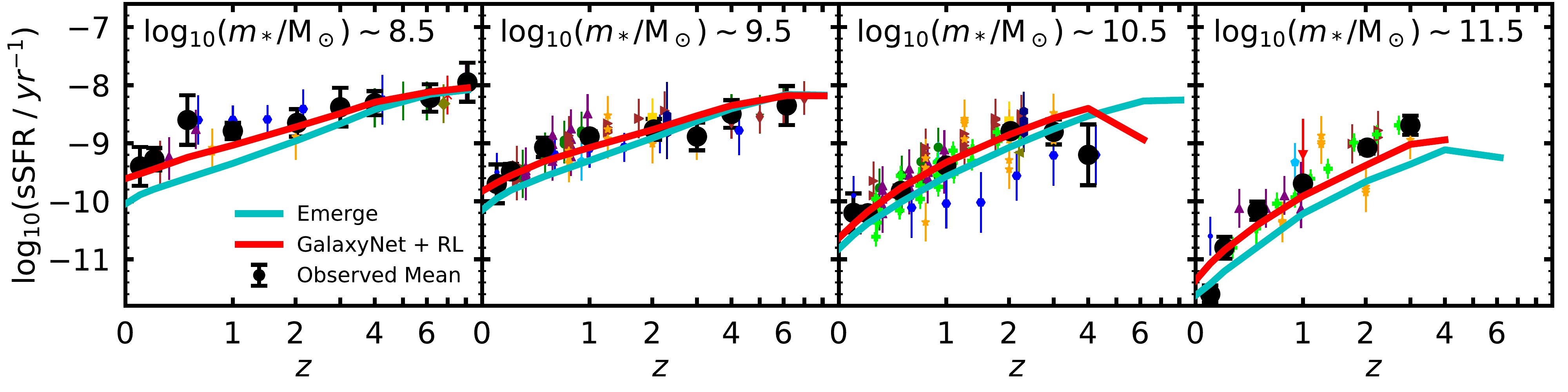}
    \caption{
        Comparison of the average specific star formation rate (sSFR) as function of redshift between \textsc{emerge} (blue lines) and GalaxyNet trained with reinforcement learning (red lines) for four stellar mass bins. The coloured symbols in the background correspond to the original observed data sets use to create the mean observed values (black symbols with error bars).
    }
    \label{fig:GalaxyNet_RL_SSFR}
\end{figure*}


\begin{figure}
	\includegraphics[width=\halfwidth]{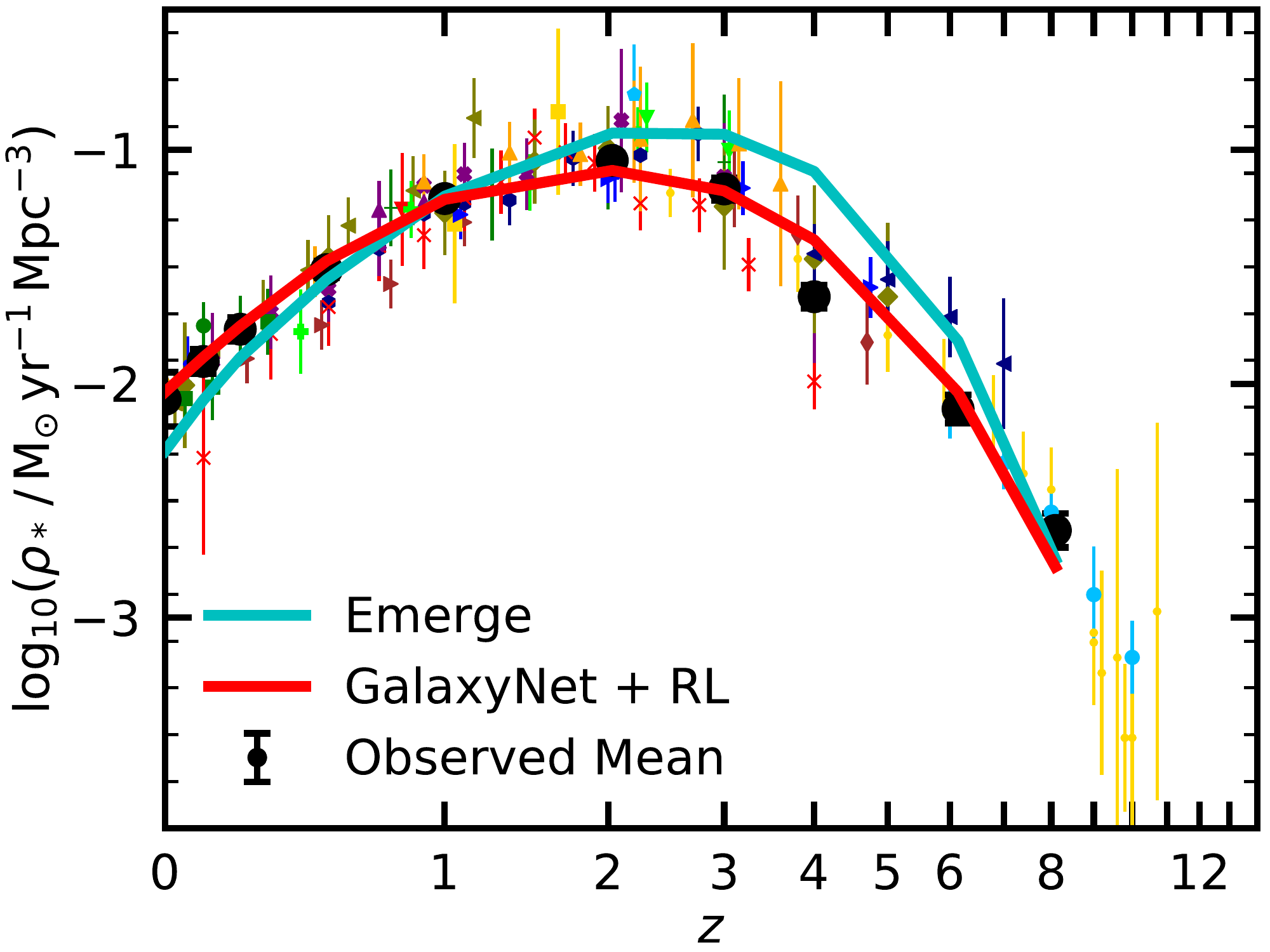}
    \caption{
        Comparison of the cosmic star formation rate density as function of redshift between \textsc{emerge} (blue lines) and GalaxyNet trained with reinforcement learning (red lines). The coloured symbols in the background correspond to the original observed data sets use to create the mean observed values (black symbols with error bars).
    }
    \label{fig:GalaxyNet_RL_CSFRD}
\end{figure}

GalaxyNet trained with reinforcement learning (GalaxyNet+RL) is able to achieve a much smaller loss value for the global statistics than the labels that were used in the supervised learning before. In Figure \ref{fig:GalaxyNet_RL_SMF} we show the SMF for all ten redshifts. The coloured symbols in the background correspond to the original observed data sets use to create the mean observed values, which are given by the black symbols with error bars. The blue lines show the results of \textsc{emerge}, which has been used for the supervised learning before. The results of GalaxyNet trained with reinforcement learning directly on the observed data are given by the red lines. At low redshift ($z\lesssim0.2$) and at high redshift ($z\gtrsim6$), both \textsc{emerge} and GalaxyNet are able to reproduce the observed SMF very well. However, at intermediate redshifts ($z=0.5 - 4$), \textsc{emerge} tends to overpredict the SMF around the knee, i.e. between $\log(m/\Msun)\sim10.5$ and 11.5, while GalaxyNet is able to fit the observations extremely well. The SMF contribution to the loss is $L_\mathrm{SMF,GN}=211$ for GalaxyNet, compared to $L_\mathrm{SMF,EM}=1353$ for \textsc{emerge}.

This result is a direct consequence of the complexity of both methods. For \textsc{emerge} the philosophy is to use as few parameters as possible to reproduce the data as best as possible, which was assessed with model selection criteria. This choice prohibited a better fit to the SMD at intermediate redshifts and masses, as this would have required more free parameters, which is penalised by the model selection criteria. GalaxyNet instead is not restricted by this issue, and uses $1,202$ free parameters to map the halo properties to the galaxy properties. Moreover, in \textsc{emerge} this mapping was parameterised by hand in a specific way (double power-law for the efficiency and linear evolution of the parameters with scale factor), while GalaxyNet finds the relation automatically. Choosing a linear scaling of the normalisation of the efficiency with scale factor leads to more star formation at high redshift and consequently to somewhat more massive galaxies with a larger SMF at intermediate redshift. We will investigate this aspect closer in section \ref{sec:efficiency}.

The fraction of quenched galaxies as a function of stellar mass is shown in Figure \ref{fig:GalaxyNet_RL_FQ}. As before, the individual observed data sets are given by the coloured symbols in the background, while the averages are given by the black symbols. The results by \textsc{emerge} (blue lines) are too high at low redshift, but fit the observations quite well at mid to high redshift. At high redshift and for massive galaxies though, they are also too high. GalaxyNet shows a good agreement with the observations at low redshift ($z<1$), but is too low for massive galaxies between $z = 1$ and 2. The quenched fraction contribution to the loss is $L_\mathrm{FQ,GN}=191$ for GalaxyNet, compared to $L_\mathrm{FQ,EM}=267$ for \textsc{emerge}. We assume that the quenched fractions cannot be reproduced to high accuracy even by GalaxyNet results from the method that has been used to compute it. While for the observations, active and passive galaxies have been divided by a cut in a colour-colour diagram, the division for GalaxyNet and \textsc{emerge} has been done with a sSFR threshold.

The projected two-point auto-correlation functions for four stellar mass bins at $z=0.1$ are shown in Figure \ref{fig:GalaxyNet_RL_WP}. Both \textsc{emerge} (blue lines) and GalaxyNet (red lines) reproduce the observed clustering (black symbols) very well. The clustering contribution to the loss is $L_\mathrm{WP,GN}=29$ for GalaxyNet, compared to $L_\mathrm{WP,EM}=34$ for \textsc{emerge}. This result is not surprising as they both reproduce the SMF at $z=0.1$ very well and consequently have a very similar SHM relation. Since the clustering is set by the dark matter haloes and mainly depends on halo mass, both models show the same clustering for galaxies.

The average sSFR as a function of stellar mass for four stellar mass bins with $\Delta \log m_* = 0.4$ dex is shown in Figure \ref{fig:GalaxyNet_RL_SSFR}. For all mass bins, GalaxyNet (red lines) reproduces the observations (black symbols) very well from low to high redshift. In contrast, \textsc{emerge} shows somewhat too low sSFRs, especially at low redshift and for high stellar masses. However, at high masses and high redshift, \textsc{emerge} shows higher sSFRs than GalaxyNet. The contribution to the loss from the sSFRs is $L_\mathrm{SSFR,GN}=34$ for GalaxyNet, while it is $L_\mathrm{SSFR,EM}=204$ for \textsc{emerge}. This result is in broad agreement with our result for the evolution of the SMF. At high redshift, galaxies in \textsc{emerge} tend to have too high sSFRs and grow to fast, leading to too many intermediate-mass galaxies at mid redshift. At low redshift, however, the sSFRs and galaxy growth are too low, such that the SMF does not grow much and is in good agreement with the observations at $z=0$ again.

The cosmic SFR density (CSFRD) as a function of redshift is shown in Figure \ref{fig:GalaxyNet_RL_CSFRD}. Similarly to our results for the sSFRs, GalaxyNet (red line) reproduces the observations (black symbols) very accurately, while \textsc{emerge} (blue line) shows too high CSFRDs at high redshift ($z\gtrsim2$) and too low CSFRDs at low redshift ($z\lesssim1$). This improved fit also becomes obvious in the loss, as the CSFRD contribution to the loss is only $L_\mathrm{CSFRD,GN}=79$ for GalaxyNet, compared to $L_\mathrm{CSFRD,EM}=362$ for \textsc{emerge}. Again we see that in \textsc{emerge}, galaxies form stars too efficiently at high redshift, so that consequently the SMF is too high at intermediate redshift, while they form slightly too few stars at low redshift, so that the SMF agrees better with the observations again.


\begin{figure*}
	\includegraphics[width=\fullwidth]{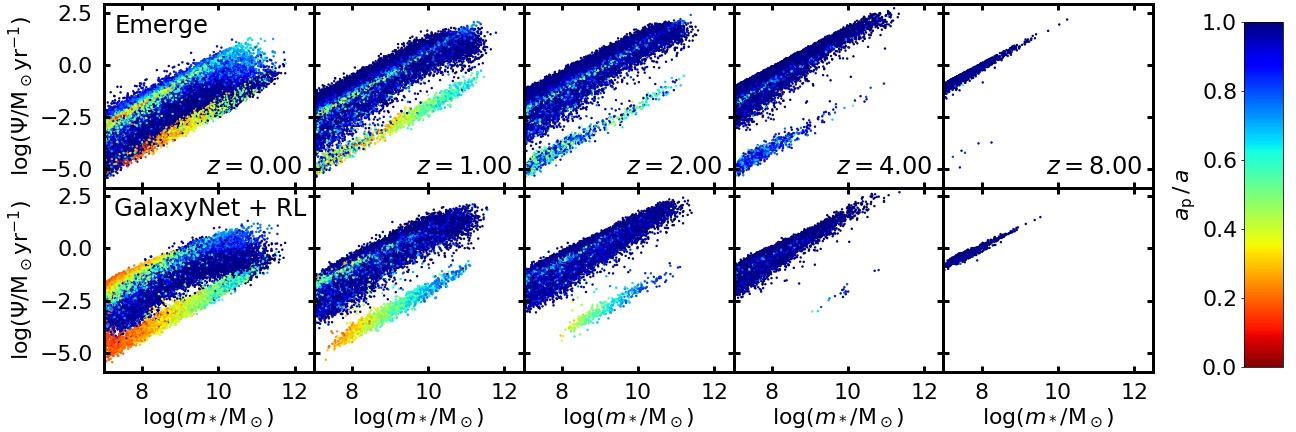}
    \caption{
        Comparison of the SFR vs stellar mass relation between \textsc{emerge} (top panels) and GalaxyNet trained with reinforcement learning (bottom panels). From left to right the panels show the results at five different redshifts. The colour of each point corresponds to the ratio between the scale factor at peak mass and the scale factor at this redshift, i.e. for blue points the halo mass peaks at the current redshift, while for red points the halo mass peaked early-on.
    }
    \label{fig:GalaxyNet_RL_MS}
\end{figure*}


\begin{figure*}
	\includegraphics[width=\fullwidth]{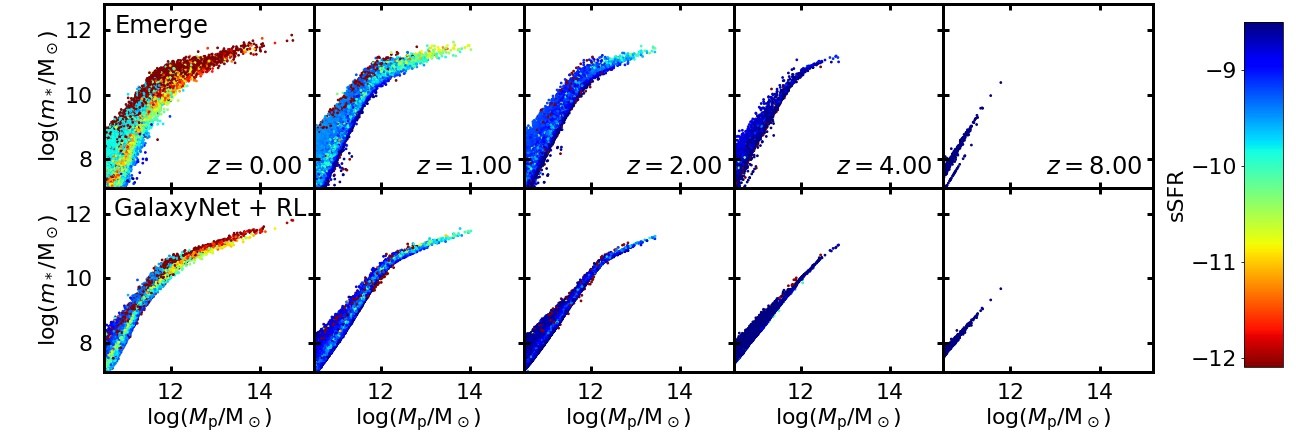}
    \caption{
        Comparison of the stellar-to-halo mass (SHM) relation between \textsc{emerge} (top panels) and GalaxyNet trained with reinfocement learning (bottom panels). From left to right the panels show the results at five different redshifts. The colour of each galaxy corresponds its sSFR. The predictions of GalaxyNet are noticeably different to those of \textsc{emerge}. At low redshift, the relation is very similar, though the scatter in GalaxyNet is considerably lower. At higher redshift, the low-mass slope is much shallower in GalaxyNet, and the normalisation is slightly lower.
    }
    \label{fig:GalaxyNet_RL_SHMR}
\end{figure*}

Having shown that GalaxyNet is able to reproduce the global statistics significantly better than \textsc{emerge} for many data sets (SMF, sSFR, CSFRD), or at an equivalent level for others (FQ. WP), we can now investigate how the two labels relate, i.e. how the SFR depends on the stellar mass. In Figure \ref{fig:GalaxyNet_RL_MS}, we show this relation for each galaxy at five different redshifts. The top and bottom panels show the results by \textsc{emerge} and GalaxyNet, respectively. The colour of the points gives the ratio $a_\mathrm{p}/a$ between the scale factor at peak mass $a_\mathrm{p}$ and the scale factor at this redshift $a$, so that for blue points the halo mass reaches its peak at the current redshift (typical for central galaxies), and for red points the halo mass peaked much earlier (typical for satellites).

In contrast to GalaxyNet trained with supervised learning (Figure \ref{fig:GalaxyNet_SL_SHMR}), GalaxyNet trained with reinforcement learning shows some unique differences to the results of \textsc{emerge}. While both models reproduce the star formation main sequence and the quenched cloud, the formation times of the haloes are different for GalaxyNet+RL. While for \textsc{emerge} at $z=0$ the haloes that peaked early (satellite galaxies) are located in the middle of the main sequence (on the ridge line), while for GalaxyNet+RL they are located at the upper end of the main sequence. Moreover, quenched low-mass galaxies in GalaxyNet+RL are almost exclusively satellites, while in \textsc{emerge} there are also many quenched low-mass centrals. Massive quenched galaxies on the other hand, are almost always living in haloes that peaked early, i.e. they are usually centrals.

At high redshift, the main difference between GalaxyNet+RL and \textsc{emerge} is the smaller amount of quenched galaxies in the former, which was also apparent in the quenched fractions (Figure \ref{fig:GalaxyNet_RL_FQ}). Further, we notice that at fixed stellar mass the SFR in GalaxyNet+RL is generally higher than the SFR in \textsc{emerge} at low mass and low redshift ($z\lesssim1$) but slightly lower at high mass and high redshift ($z\gtrsim2$), in agreement with the results for the sSFRs (Figure \ref{fig:GalaxyNet_RL_SSFR}). Finally, we observe a slightly shallower slope of the main sequence for GalaxyNet+RL. 

The relation between peak halo mass and stellar mass is shown in Figure \ref{fig:GalaxyNet_RL_SHMR} at five redshifts for \textsc{emerge} (upper panels) and GalaxyNet trained with reinforcement learning (lower panels). The colour of each point corresponds to the sSFR of each galaxy. At low redshift ($z<1$), both models agree fairly well, although GalaxyNet has considerably less scatter at fixed halo mass. This is not surprising, as both models reproduce the local SMF very well. At higher redshift, however, the SHM relation for GalaxyNet+RL is significantly different from \textsc{emerge}. First, the low-mass slope ($\beta$) is much shallower. Second, the normalisation is lower in GalaxyNet+RL. Consequently, the stellar mass of galaxies in intermediate-mass haloes is much lower. For example, at $z=2$ the average stellar mass of a galaxy in a halo with $\log(M_\mathrm{p}/\Msun)=12$ is only $\log(m_*/\Msun)=10.0$, i.e. a factor 2 lower compared to \textsc{emerge} with $\log(m_*/\Msun)=10.3$. This result is a direct consequence of the different fits to the SMF at intermediate redshift. While the number of intermediate-mass galaxies in \textsc{emerge} can be too high by up to a factor of 2 compared to the observations, GalaxyNet+RL reproduces the data very accurately. Consequently, the SHM relation given by GalaxyNet is significantly more reliable.


\begin{table}
	\centering
	\caption{The parameters of the stellar-to-halo mass (SHM) ratio given by a double-power-law (eqn. \ref{eqn:epsilon}) at different redshifts as found by GalaxyNet+RL.}
	\label{tab:shmr}
	\begin{tabular}{ccccc}
		\hline
		$z$ & $M_1$ & $\epsilon_\mathrm{N}$ & $\beta$ & $\gamma$\\
		\hline
		0.0 & $11.94\pm0.04$ & $0.116\pm0.003$ & $1.26\pm0.06$ & $0.52\pm0.02$\\
		0.1 & $11.97\pm0.03$ & $0.116\pm0.003$ & $1.25\pm0.05$ & $0.52\pm0.02$\\
		0.2 & $11.99\pm0.02$ & $0.115\pm0.002$ & $1.23\pm0.03$ & $0.53\pm0.01$\\
		0.5 & $12.05\pm0.02$ & $0.111\pm0.001$ & $1.21\pm0.03$ & $0.53\pm0.01$\\
		1.0 & $12.15\pm0.02$ & $0.109\pm0.002$ & $1.14\pm0.04$ & $0.56\pm0.01$\\
		2.0 & $12.31\pm0.04$ & $0.097\pm0.002$ & $1.01\pm0.04$ & $0.58\pm0.03$\\
		3.0 & $12.51\pm0.03$ & $0.096\pm0.001$ & $0.84\pm0.02$ & $0.61\pm0.03$\\
		4.0 & $12.55\pm0.06$ & $0.095\pm0.002$ & $0.75\pm0.03$ & $0.62\pm0.11$\\
		6.0 & $12.61\pm0.10$ & $0.062\pm0.001$ & $0.63\pm0.02$ & $0.64\pm0.19$\\		
		\hline
	\end{tabular}
\end{table}

At a given redshift, the SHM ratio, i.e. the integrated conversion efficiency, can be parameterised as a double-power-law, as given by eqn. \ref{eqn:epsilon}, where $\epsilon(M,z)$ corresponds to the ratio between stellar and total baryonic mass (the halo mass times the universal baryonic fraction), $m_*/m_\mathrm{b} = m_*/(f_\mathrm{b} M_\mathrm{h})$. The parameters of this relation for different fixed redshifts are presented in Table \ref{tab:shmr}. The characteristic halo mass where the ratio peaks ($M_1$) evolves more strongly towards high redshift than predicted by \textsc{emerge}, and the low-mass slope ($\beta$) is shallower, while the high-mass slope is very similar. In contrast to the empirical model, the normalisation ($\epsilon_\mathrm{N}$) decreases with increasing redshift.


 \begin{figure*}
	\includegraphics[width=\fullwidth]{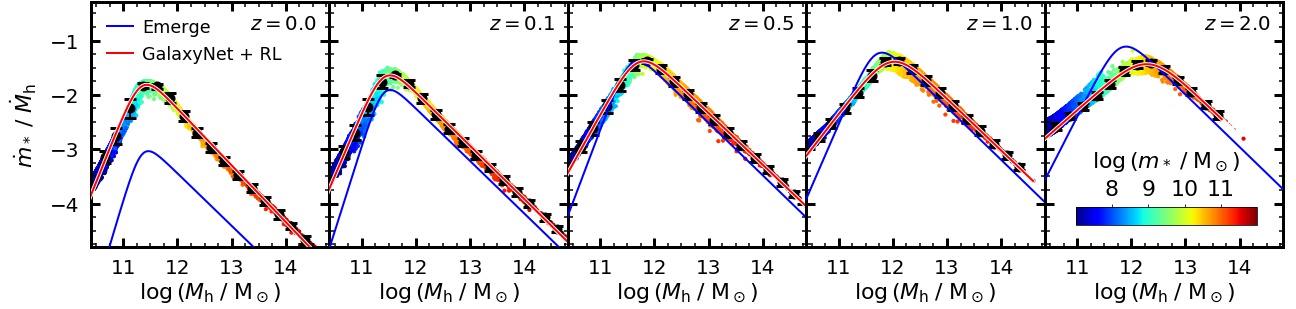}
    \caption{
        The instantaneous baryon conversion efficiency (ratio between SFR of a galaxy and the growth rate of its halo) for systems that are currently at their peak halo mass (central galaxies). The colour of each point indicates the stellar mass of each galaxy as given by the colour bar. The individual points show the results for each galaxy by GalaxyNet trained with reinforcement learning, the black symbols give the average at fixed peak halo mass, and the red lines represent double-power-law fits. The blue lines indicate the fits that are applied in \textsc{emerge}.
    }
    \label{fig:efficiency}
\end{figure*}

We generally find less scatter in stellar mass at fixed halo mass in GalaxyNet+RL compared to \textsc{emerge}. The reason for this is likely the connection between the two targets, the stellar mass and the SFR. While in \textsc{emerge} these two quantities are related self-consistently, i.e. the stellar mass is the time integral of the SFR (modulo mass loss), in GalaxyNet the there is no similar hard constraint between them. Thus, in \textsc{emerge} the stellar mass strongly depends on the star formation history of the galaxy, such that two haloes of the same (peak) mass can have galaxies with very different stellar masses. In GalaxyNet, however, the stellar mass depends much less on the individual formation history (only on the peak mass and the corresponding scale factor and growth rate), so that the range of stellar masses at a fixed halo mass is much smaller.

Now that we have shown that GalaxyNet trained with reinforcement learning is able to predict galaxy properties such as the stellar mass and SFR and reproduce observed global statistics very accurately, we can use the model to study the evolution of galaxies in more detail, and predict several quantities that have not been used to train the model. In the following two sections, we will investigate how the baryon conversion efficiency evolves with redshift and halo mass, and apply GalaxyNet to one of the largest cosmological simulations to predict galaxy clustering up to high redshift.


\section{Predictions for the conversion efficiency}
\label{sec:efficiency}


 \begin{figure}
	\includegraphics[width=\columnwidth]{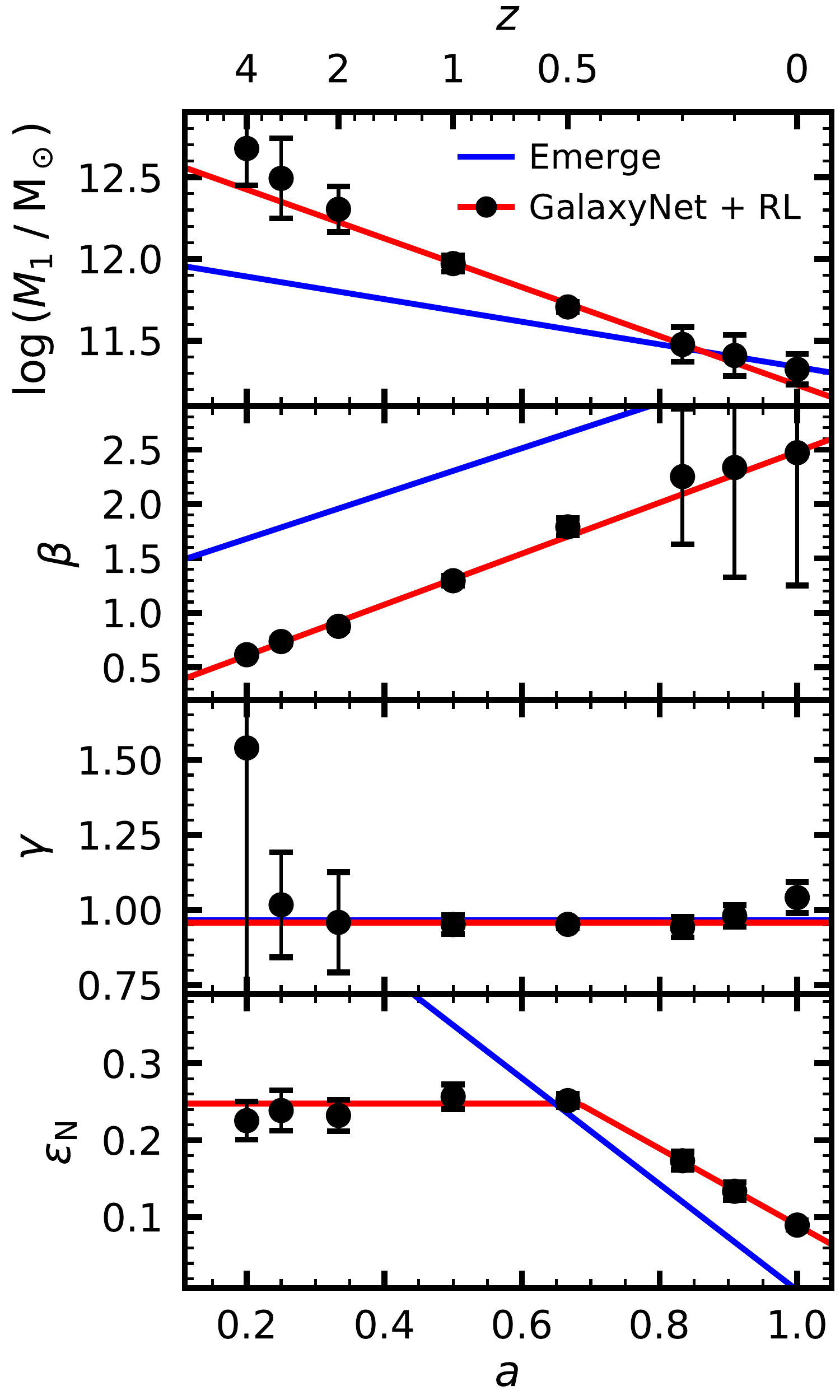}
    \caption{
        The redshift dependence of the parameters of the instantaneous baryon conversion efficiency. The four parameters correspond to those of a double-power-law as given by eqn. \ref{eqn:epsilon}. The symbols with error bars have been derived by fitting a double-power-law to the results of GalaxyNet+RL at different redshifts (Figure \ref{fig:efficiency}). The red lines show fits to the black symbols, and the blue lines show the relations that have been used in \textsc{emerge}.
    }
    \label{fig:parameters}
\end{figure}

Similar to other empirical galaxy formation models, GalaxyNet populates dark matter haloes with galaxies, i.e. it maps the properties of haloes to the properties of galaxies. The big advantage of GalaxyNet is that it is able to perform this mapping in a way that reproduces observed statistics far more accurately than any other current model. Although as a neural network, GalaxyNet is not directly interpretable, such that its parameters could give insight into the physics of galaxy formation, we can employ GalaxyNet to investigate how the connection between halo and galaxy properties evolves through cosmic time. Once this connection has been established, we can then parameterise it and infer the evolution more directly. Further, a parameterised relation can be easily included in an empirical galaxy formation method that follows individual haloes through time in a self-consistent manner.

To this end we compute the instantaneous baryon conversion efficiency for each system, which is given by the ratio between the SFR of the galaxy $\dot m_*$ and the mass growth rate of the associated dark matter halo $\dot M$, and describes how efficiently the infalling baryons ($f_\mathrm{b}\dot M$) are converted into stars at the centre. As this quantity is only sensibly defined for haloes that still grow, i.e. for central galaxies, we compute it only for systems that are currently at their peak halo mass. In this respect, the computed efficiency closely follows the one parameterised in \textsc{emerge}, which also only applies as long as the halo grows. We plot the computed instantaneous baryon conversion efficiency as a function of peak halo mass in Figure \ref{fig:efficiency} for five redshifts. The colour of the individual points correspond to the stellar mass of each galaxy as given by the colour bar. The black symbols with error bars give the average conversion efficiency at fixed halo mass, while the red lines show a double-power-law fit. In comparison, we show the fits that have been used in \textsc{emerge} (blue lines). At fixed redshift, the relation is very tight with little scatter at a given halo mass. The normalisation increases towards higher redshift, but reaches its maximum around $z\sim0.5$. The characteristic mass evolves strongly with redshift. While the high-mass slope does not evolve at all, the low-mass slope is very steep at low redshift, but much shallower at high redshift. In \textsc{emerge}, the normalisation increases much more strongly, while the characteristic mass evolves considerably less. Moreover, the low-mass slope in \textsc{emerge} is generally steeper, but the high-mass slope is very similar.

To investigate the redshift evolution in more detail, we determine how each parameter of the double-power-law evolves with redshift. We fit the instantaneous baryon conversion efficiency at each redshift, and plot the resulting parameters In Figure \ref{fig:parameters} (black symbols with error bars). Interestingly, almost all parameters have a linear dependence on the scale factor $a$ with very little scatter, as has been assumed for \textsc{emerge}. The characteristic mass decreases linearly with $a$, the low-mass slope increases linearly with $a$, and the high-mass slope is constant. However, the only deviation from this linear trend is the normalisation. We find that at high redshift it is almost constant, but then at $z\sim0.5$ the normalisation starts to decrease linearly with increasing scale factor. Due to this scaling, GalaxyNet is able to reproduce the observed SMF more accurately than \textsc{emerge}. At high redshift, the normalisation is low enough such that the resulting SMF is not too high, while at lower redshift the normalisation is increased leading to a stronger growth of the SMF.

As a reference, we show the scalings of the parameters for the instantaneous conversion efficiency that are used in \textsc{emerge} (blue lines). The linear scaling with scale factor has been assumed for all parameters except for the high-mass slope $\gamma$ which has been set to a constant. The values for this scaling have been determined by fitting the \textsc{emerge} predictions to the same observed statistics in \citet{Moster:2018aa}. We find that the characteristic mass $M_1$ evolves more strongly with redshift in GalaxyNet ($>1$ dex) compared to \textsc{emerge} ($\sim0.5$ dex). The low-mass slope $\beta$ decreases with increasing redshift in a very similar way in both models, however, in GalaxyNet the slope is much shallower at all redshifts compared to \textsc{emerge}, with an offset of about -1. The high-mass slope $\gamma$ is found to be constant in GalaxyNet as well, with a value that is very close to the one found in \textsc{emerge}. The only major difference between the two models is the normalisation $\epsilon_\mathrm{N}$, which was assumed to follow a linear relation with $a$ in \textsc{emerge}, but is found to stay constant at high redshift and then to decrease linearly with $a$ in GalaxyNet. As a consequence of assuming a simple linear relation in \textsc{emerge}, star formation at high redshift is too strong and the resulting SMF at intermediate redshift is somewhat too high. At low redshift, star formation in \textsc{emerge} is lower than in GalaxyNet, so that the SMF grows only relatively little and is again in very good agreement with the observations at $z=0$. In future work, we will implement this different parameterisation for $\epsilon_\mathrm{N}(z)$ in \textsc{emerge} in an effort to get an even better fit to the SMFs.

For GalaxyNet+RL, we find the following redshift scaling relations for the double-power-law parameters that describe the instantaneous baryon conversion efficiency (cf. eqn. \ref{eqn:epsilon}):
\begin{align}
\log_{10} M_1(z)& = 11.23 + 1.50\; z\,/\,(z+1) \; \notag \\[0.5\baselineskip]
\epsilon_\mathrm{N}(z)& = 0.09 + 0.50\; z\,/\,(z+1) \; \notag \\[0.5\baselineskip]
\beta(z)& = 2.48 - 2.34\; z\,/\,(z+1) \; \notag \\[0.5\baselineskip]
\gamma(z)& = 0.96 \notag
\end{align}


\section{Predictions for large scale structure}
\label{sec:lss}


As we have shown in section \ref{sec:reinforcement}, GalaxyNet trained with reinforcement learning is able to reproduce observed statistics very accurately. Therefore it is the ideal model to populate $N$-body simulations with galaxies and make predictions on galaxy properties and statistics that have not been used in the training. Of course, the model can only be applied in the feature range in which has been trained, as machine learning models tend to be very unstable when extrapolated beyond this range. One primary use case of this approach is predicting galaxy clustering up to high redshift. As the clustering properties on large scales are governed by the clustering of dark matter haloes, and not by the formation of galaxies inside of the haloes, we can apply the trained GalaxyNet to a much larger cosmological volume as long as the feature values, i.e. halo properties, are not extrapolated beyond the training range. Consequently, we exclude galaxies in the most massive haloes from our analysis.

This technique has large advantages over models that trace the formation of galaxies in individual haloes from high redshift, such as empirical models (e.g. \textsc{emerge}, UniverseMachine, Steel), semi-analytic models, and hydrodynamic simulations. In simulations with large cosmological volumes they achievable resolution is typically quite low, so that the low-mass haloes (below the characteristic mass $M_1$) are not resolved and cannot be identified, which means that low-mass galaxies cannot be modelled. However, as galaxies in more massive haloes have a considerable ex-situ growth, but the low-mass galaxies are not present in the model, massive galaxies cannot grow sufficiently, such that it is not possible to get their properties correctly. It is therefore much simpler to apply the results of a model that does not explicitly trace haloes through time, but that simply relies on the halo properties at the selected redshift.

 \begin{figure}
	\includegraphics[width=\halfwidth]{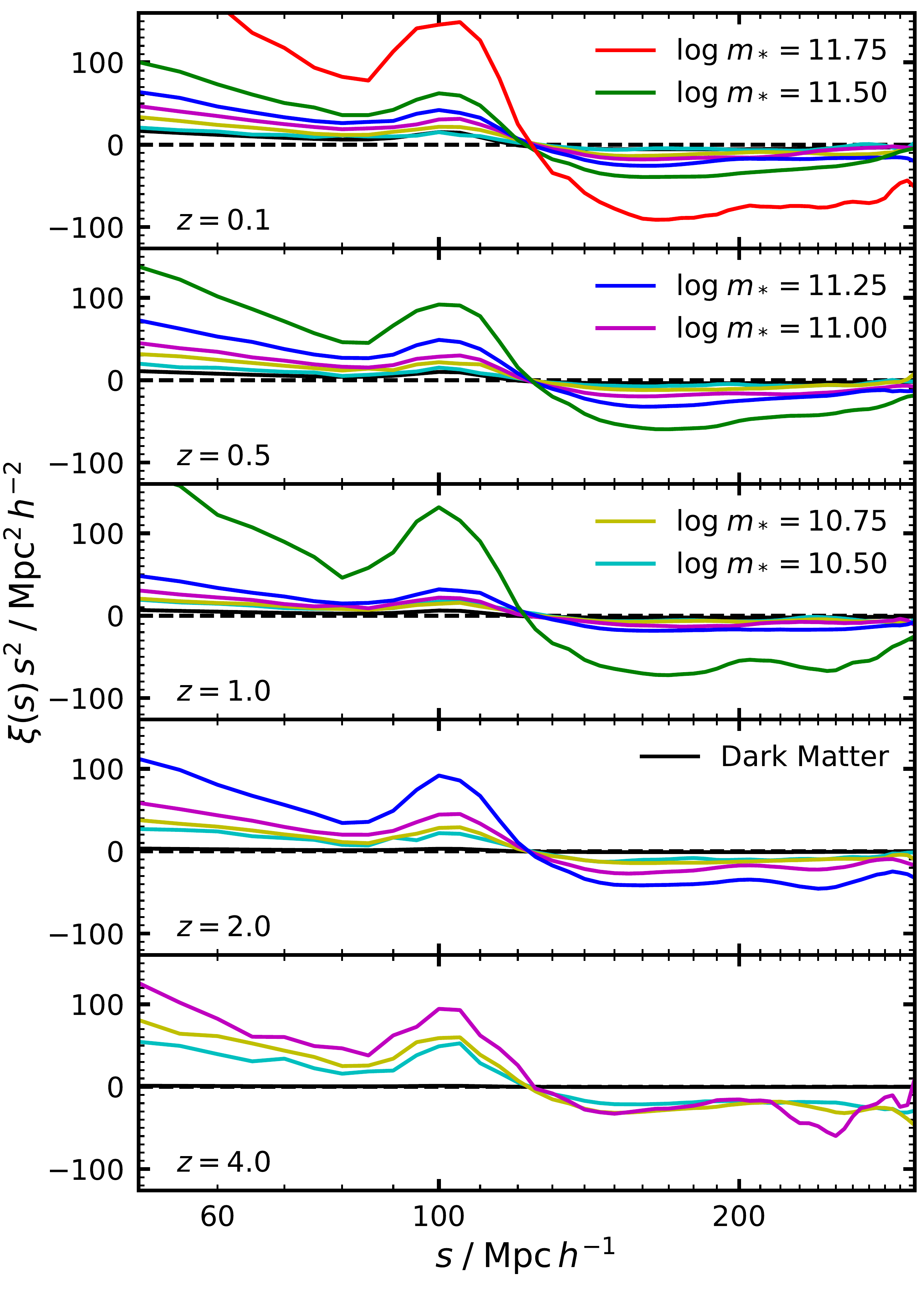}
    \caption{
        Redshift-space correlation functions $\xi(s)$ at four different redshifts for galaxies in several stellar mass bins (coloured lines) derived from applying GalaxyNet trained with reinforcement learning to the Huge MultiDark Planck simulation. The real-space linear correlation functions for the dark matter density field (black lines) are included for comparison. The peak of the BAO signal is at slightly smaller scales for massive galaxies, while the zero-crossing happens at the same scale of $s_0 = 123\Mpc\,h^{-1}$ (within 1 per cent), independent of stellar mass and redshift.
    }
    \label{fig:xis_bao}
\end{figure}

To be able to predict galaxy clustering up to large scales, we employ one of the largest $N$-body simulations available to date, the Huge MultiDark Planck simulation \citep[HugeMDPL;][]{Klypin:2016aa}, which belongs to a series of MultiDark simulations using the same Planck cosmology as quoted in section \ref{sec:features}. The simulation box has a side length of $4 \, h^{-1}\Gpc = 5.9\Gpc$, and contains $4096^3$ collisionless particles, which corresponds to a particle mass of $1.2\times10^{11}\Msun$. The initial conditions were evolved with the {\sc Gadget2} code using a gravitational softening of $37\kpc$ from $z=100$ to 0, saving 103 snapshots. Dark matter haloes and subhaloes were identified with the {\sc Rockstar} halo finder, and halo merger trees were generated with the {\sc ConsistentTrees} code. From these merger trees we created halo catalogues including orphans at 10 snapshots, containing the same halo properties as used before (see Table \ref{tab:features}).

We apply the same feature scaling as used before, and use GalaxyNet trained with reinforcement learning to derive the stellar mass and SFR of the corresponding galaxy. Because of the resolution limits of the simulation, its HMF deviates from the HMF of simulations with higher resolution below a halo mass of $\log(M/\Msun)=12.3$, which can therefore be seen as the minimum halo mass. The corresponding average stellar mass is $\log(m/\Msun)=10.5$, and the upper bound given the scatter is $\log(m/\Msun)=10.6$, so that all stellar mass bins above this value will not be affected by unresolved haloes. Similarly, at the massive end, we trained GalaxyNet in haloes up to $\log(M/\Msun)=15.1$, which corresponds to an average stellar mass of $\log(m/\Msun)=12.0$ and a lower bound given the scatter of $\log(m/\Msun)=11.9$, which means that up to this mass the model is applicable. Within these stellar mass limits we will now use the galaxy catalogues created with GalaxyNet+RL to study galaxy clustering properties and to make predictions for future surveys. We focus on the stellar mass dependence of the baryonic acoustic oscillation (BAO) signal, the galaxy bias, and the projected correlation functions for active and passive galaxies.


\subsection{The stellar mass dependence of the BAO signal}

Acoustic density waves in the primordial plasma of the early universe led to fluctuations in the density of the baryonic matter that were frozen in at the epoch of recombination. The evolution of these baryonic acoustic oscillations (BAO) imprints a distinct length scale on the galaxy distribution, which corresponds to the maximum distance the acoustic waves could travel in the primordial plasma before recombination. This provides a `standard ruler' for distance estimates, so that the BAO scale can be used to constrain the cosmology and the nature of dark energy. Galaxy surveys, such as BOSS \citep{Ross:2017aa} and DES \citep{Abbott:2019aa} have measured the BAO signal in the low-redshift ($z \lesssim 1$), and future surveys, such as LSST and Euclid are pushing these measurements to higher redshift ($z\sim 2$). Since these observations can only measure biased tracers of the density field, it is crucial to have theoretical models that predict the galaxy distribution for a specific cosmology without depending on physical processes that are not well understood such as feedback from baryons. Therefore empirical models and models that are based on machine learning are the ideal tools to leverage these cosmological probes.

It is well known that the clustering of galaxies with respect to the dark matter density field, i.e. the galaxy bias, depends on the stellar mass of the galaxies. Here, we investigate how the BAO signal depends on stellar mass at five different redshifts based on GalaxyNet trained with reinforcement learning. To this end we populate all dark matter haloes in the HugeMDPL simulation with galaxies using GalaxyNet+RL to get their stellar mass and SFR. We then take the locations of the dark matter halo centres as the galaxy positions and transform them to redshift space using HaloTools, which computes the redshift distortions caused by each galaxy's peculiar velocity. We then divide the galaxies at the five redshifts $z = 0.1$, 0.5, 1.0, 2.0, and 4.0 into stellar mass bins, and compute the redshift-space two-point correlation function $\xi(s)$ for each sample with HaloTools. We show the results in Figure \ref{fig:xis_bao}, where each panel corresponds to one redshift, and each of the coloured lines corresponds to one stellar mass bin, with the mean stellar mass given in the legend. The corresponding real-space linear correlation function $\xi(r)$ (black lines) is included for comparison in each panel.

We find that the peak of the BAO signal slightly depends on stellar mass at all redshifts, with a peak at smaller scales for more massive galaxies. However, at all redshifts, the zero-crossing of $\xi(s)$, which is a consequence of the power spectrum approaching 0 at large scales, happens at the same (comoving) length scale of $s_0 = 123\Mpc\,h^{-1}$ (within 1 per cent), independent of stellar mass and redshift. A similar finding was obtained by \citet{Prada:2011aa} for haloes of fixed bias. This independence of mass and redshift makes the zero-crossing scale $s_0$ an ideal candidate to probe cosmology.


\subsection{The stellar mass dependence of the bias}

\begin{figure}
	\includegraphics[width=\columnwidth]{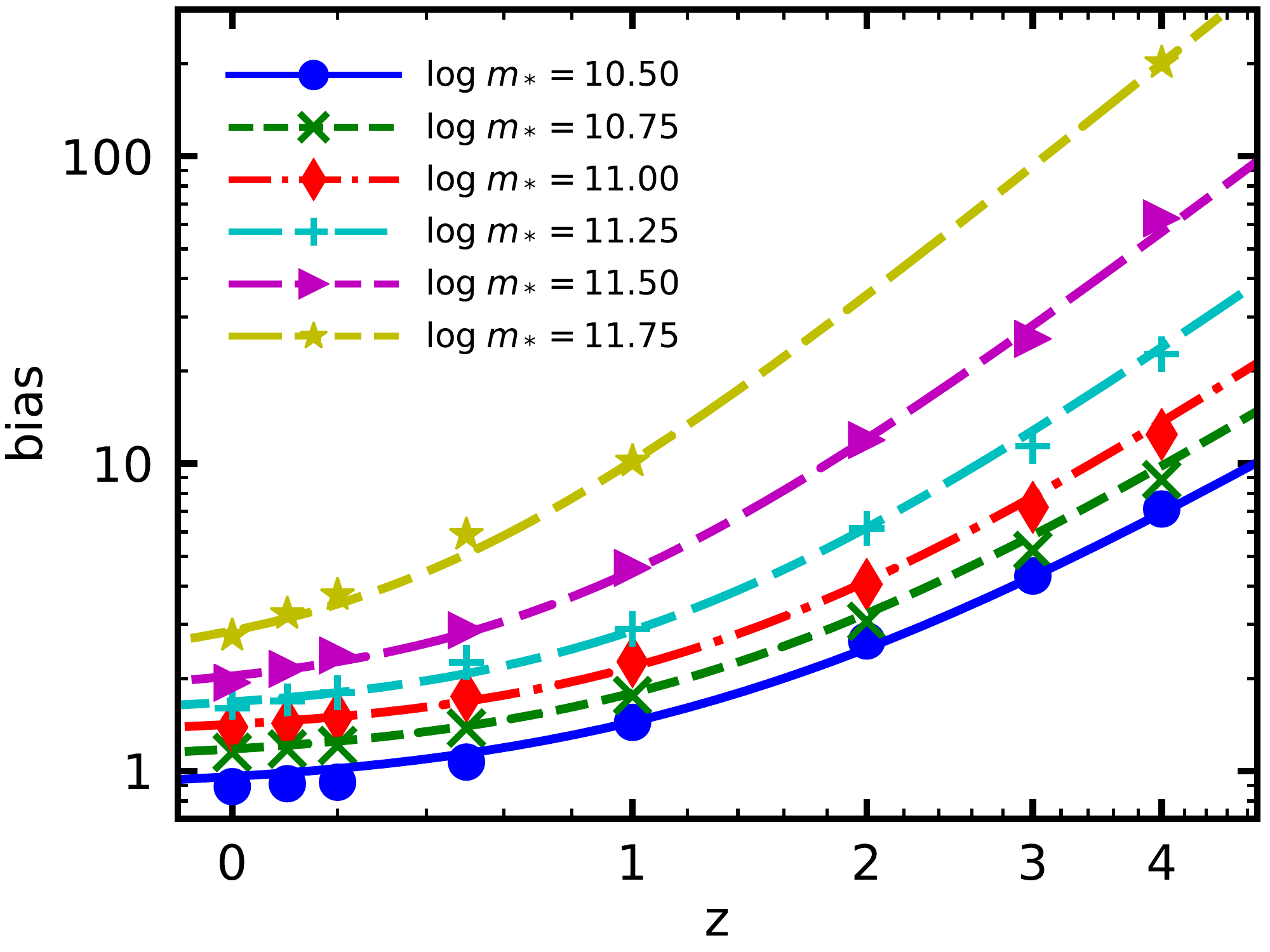}
    \caption{
        Galaxy bias at a fixed scale ($\sim 8\Mpc\,h^{-1}$) as a function of redshift for different stellar masses. The symbols have been derived by averaging the bias over a distance interval while the lines are fits to the symbols (using eqn. \ref{eqn:bias} with parameters in Table \ref{tab:bias}).
    }
    \label{fig:bias}
\end{figure}

The clustering amplitude of galaxies with respect to the clustering of the dark matter can be expressed as the galaxy bias, which is defined as the square root of the ratio between the two-point correlation function of a galaxy sample $\xi_\mathrm{gg}(r,m,z)$ and the correlation function of the dark matter $\xi_\mathrm{dm}(r,z)$, i.e. $b = \sqrt{\xi_\mathrm{gg}/\xi_\mathrm{dm}}$. We use the real-space galaxy correlation functions predicted with GalaxyNet+RL for 6 stellar mass bins at 8 redshifts and compute the bias by averaging between 4 and $16\Mpc\,h^{-1}$, where we find $b(r)$ to be almost constant. The resulting redshift evolution of the bias is shown in Figure \ref{fig:bias} for all stellar mass bins (symbols).

The galaxy bias can be approximated very well with an offset power-law fitting function
\begin{equation}
\label{eqn:bias}
b(z) = b_0 \, (z+1)^{b_1} + b_2
\end{equation}
with the parameters $b_0$, $b_1$, and $b_2$ depending on the stellar mass as given in Table \ref{tab:bias}. As found by previous studies \citep[e.g.][]{White:2007aa,Brown:2008aa,Moster:2010aa} the bias increases with redshift, although our results considerably extend the maximum stellar mass and redshift, and improve the accuracy. Massive galaxies are clustered more strongly than their low-mass counterparts, and the redshift evolution is more rapid in massive galaxies. As this behaviour is also observed for dark matter haloes, our result is a direct consequence of the monotonically increasing SHM relation.

\begin{table}
	\centering
	\caption{
	Parameters for the redshift evolution of the galaxy bias (eqn. \ref{eqn:bias}), where each line gives the parameters for galaxies with a stellar mass as given in the first column.
	}
	\label{tab:bias}
	\begin{tabular}{cccc}
		\hline
		$\log(m_*/\Msun)$ & $b_0$ & $b_1$ & $b_2$\\
		\hline
		10.50 & $0.10\pm0.01$ & $2.55\pm0.08$ & $0.86\pm0.05$\\
		10.75 & $0.11\pm0.01$ & $2.72\pm0.06$ & $1.07\pm0.03$\\
		11.00 & $0.12\pm0.01$ & $2.88\pm0.05$ & $1.30\pm0.04$\\
		11.25 & $0.16\pm0.03$ & $3.07\pm0.14$ & $1.52\pm0.18$\\
		11.50 & $0.28\pm0.03$ & $3.28\pm0.05$ & $1.76\pm0.07$\\
		11.75 & $0.71\pm0.09$ & $3.50\pm0.08$ & $2.15\pm0.30$\\
		\hline
	\end{tabular}
\end{table}

Providing a fitting function for the galaxy bias can be very useful to easily make predictions for the distribution of galaxies, e.g. to create mock galaxy catalogues, as the dark matter correlation function can be computed analytically and the multiplied with the squared bias. Similarly, the bias can be used in conjunction with the simple method to determine the cosmic variance for any survey design as described by \citet{Moster:2011aa}. It provides analytic functions to determine the cosmic variance of dark matter given a pencil beam geometry and redshift bin size, which can then be multiply with the bias presented here to get the cosmic variance for galaxies of a particular stellar mass. In this way it becomes possible to determine the uncertainty of measurements by galaxy surveys.


\subsection{The clustering of active and passive galaxies}

 \begin{figure*}
	\includegraphics[width=\fullwidth]{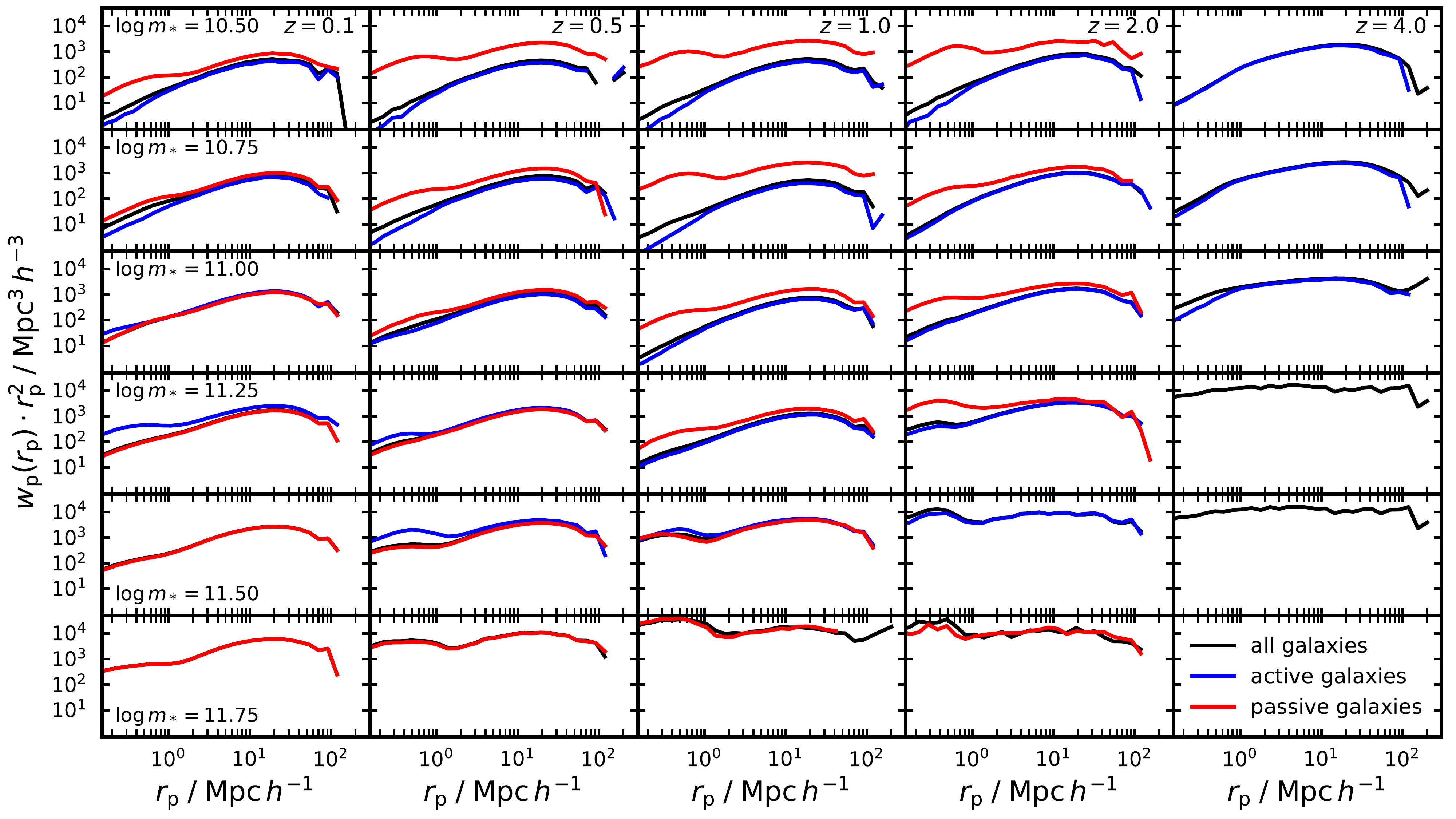}
    \caption{
        Projected galaxy correlation functions $w_\mathrm{p}(r_\mathrm{p})$ derived with GalaxyNet+RL for all galaxies at five different redshifts (columns) and several stellar mass bins (rows). The results for all, active, and passive galaxies are given by the black, blue, and red lines, respectively. Passive galaxies cluster more strongly than active galaxies, except for massive galaxies at low redshift.
    }
    \label{fig:wp}
\end{figure*}

Dark matter haloes of a specific mass do not all form at the same time, but have a variety of formation histories. The clustering of dark matter haloes has been shown to not only depend on mass, but also on its assembly history, i.e. on properties such as the formation age or the halo concentration \citep{Gao:2005aa}, which has been named `halo assembly bias'. Similarly galaxies of a specific stellar mass that form early are more strongly clustered than galaxies that form most of their stars at late stage \citep[e.g.][]{Zehavi:2005aa}, which is often called `galaxy assembly bias'. Theoretical models found that galaxy assembly bias is a consequence of halo assembly bias if the star formation activity is linked to the halo growth history \citep[e.g.][]{Hearin:2013aa,Moster:2018aa}. Here, we investigate the clustering properties of the most massive galaxies as derived from GalaxyNet+RL, and provide predictions for missions like Euclid.

To this end we divide our galaxies in active and passive samples as described in section {sec:reinforcement}, and compute the projected two-point auto-correlation functions for each with HaloTools for 6 stellar mass bins at 5 redshifts. The results are presented in Figure \ref{fig:wp} for all galaxies (black lines), active galaxies (blue lines), and passive galaxies (red lines). We find the well-know trends, such as an increased clustering amplitude for more massive galaxies and higher redshift. For intermediate-mass galaxies with $\log(m_*/\Msun) \lesssim 11$, we recover that passive galaxies cluster more strongly than active galaxies. However, we also find that GalaxyNet+RL predicts that this trend inverts for the most massive galaxies, such that very massive active galaxies have stronger clustering than their passive analogues, especially in the one-halo term. This trend is stronger at low redshift and decreases towards higher redshift.

We believe that this effect is caused by the rarity of active galaxies among the most massive samples. While it is possible that GalaxyNet+RL is not able to recover the properties of these rare galaxies well, we argue that these rare active galaxies are all located in very few regions of increased halo growth, while the majority of massive galaxies are distributed somewhat more evenly in the more common environments where haloes do not grow anymore. This would further explain why the effect vanishes at higher redshift ($z\gtrsim1$), as active galaxies and regions with strong halo growth become more common such that they are distributed more evenly reducing the clustering amplitude. Indeed, at higher redshift passive galaxies become rarer, and are located in very few regions resulting in stronger clustering. Ultimately, this prediction needs to be tested by next-generation observations.


\section{Summary and conclusions}
\label{sec:conclusions}

In this paper, we connected the properties of dark matter haloes and galaxies through a wide \& deep neural network, called GalaxyNet. To this end, we first selected the most important features given by the properties of dark matter haloes extracted from a numerical cosmological simulation, that were needed to reproduce the targets given by the stellar mass and SFR predicted by the empirical galaxy formation model \textsc{emerge}. This feature selection was done by performing a feature importance analysis with random forests, which showed that by far the most important feature with 64 per cent relative importance is peak halo mass, i.e. the maximum virial mass of the dark matter halo through its history. This was followed by the average halo growth rate within one dynamical halo time before the halo mass peaked (10 per cent), the scale factor of this peak (10 per cent), and the scale factor at observation (6 per cent). Other features, such as the half-mass scale factor or the spin were found to be of little importance to reproduce the empirically constrained stellar masses and SFRs.

We proceeded to connect halo and galaxy properties with a neural network using supervised learning. Here, we opted for a wide \& deep architecture, which contains a deep part (regular MLP) and a wide part that connects the features directly to the output layer, and therefore allows the network to learn both deep patterns (deep path) and simple rules (wide path). For our standard number of features of 8, the deep network contains 4 hidden fully connected layers with 16 nodes, followed by 2 hidden layers with 8 nodes, and an output layer with 2 targets (Figure \ref{fig:WDNN}). We used SELUs as the activation functions and implemented GalaxyNet in TensorFlow 2. GalaxyNet was first trained with supervised learning on the galaxy properties predicted by \textsc{emerge} using the standard backpropagation algorithm with mean absolute error as loss function. We found that GalaxyNet+SL was able to reproduce the labels provided by \textsc{emerge} very accurately, with a mean absolute error of 1.25 per cent. Consequently, derived relations, such as the relation between SFR and stellar mass or the SHM relation at different redshifts were found to be identical for \textsc{emerge} and GalaxyNet+SL.

Training a machine learning model with supervised learning requires labeled data, which for the problem of connecting halo to galaxy properties can only be obtained by invoking some model. The disadvantage is that the machine learning model can at best only reproduce the invoked model, which firstly may have its own problems, and secondly is already available. Therefore we trained GalaxyNet directly on observed statistical data applying a reinforcement learning approach. To this end, we first computed the properties of all galaxies in the cosmological volume with fixed parameters (network weights and biases) and then compared the derived model statistics to the observations to get the loss. We used stellar mass functions, quenched fractions, cosmic and specific SFRs, and clustering for this comparison. The parameters were then adjusted to minimise the loss function using particle swarm optimisation invoking 50 particles. We found the the final loss after 60 steps to be significantly lower than the loss obtained with the \textsc{emerge} data, with $L_\mathrm{PSO} = 544$ vs $L_\mathrm{EM} = 2220$. We proceeded to analyse the results of GalaxyNet trained with reinforcement learning and derived the following conclusions:

(i) \textit{The stellar-to-halo mass relation}: We found that at high redshift ($z \gtrsim 1$), GalaxyNet+RL predicts a different SHM relation compared to \textsc{emerge} due to the improved fit to the observed SMFs. While at low redshift both reproduced the SMF very well, and thus have the same SHM relation, the normalisation at higher redshift is considerably lower in GalaxyNet+RL, both for the SMF and the SHM relation. Similarly, we found a shallower slope of the SHM relation at high redshift compared to \textsc{emerge}. We found that the stellar mass in haloes with $\log(M_\mathrm{p}/\Msun)=12$ is $\log(m_*/\Msun)=10.0$, a factor 2 lower than the \textsc{emerge} result. We provided fitting functions for the SHM relation found by GalaxyNet+RL.

(ii) \textit{The relation between SFR and stellar mass}: While GalaxyNet+RL shows the same relation as \textsc{emerge}, both in good agreement with the observational constraints, we found a different relation to the formation times of their haloes. We found that galaxies in haloes that reached their peak mass early (satellites) are located at the upper end of the star formation main sequence. Further, quenched low-mass galaxies in GalaxyNet+RL are almost exclusively satellites. On the other hand, the majority of massive quenched galaxies are centrals. Overall, we found a slightly shallower slope for the main sequence compared to \textsc{emerge}.

(iii) \textit{The instantaneous baryon conversion efficiency}: We used the predictions by GalaxyNet+RL to investigate the redshift evolution of the instantaneous conversion efficiency, i.e. the relation between SFR and halo growth rate. While in \textsc{emerge} a linear dependence on the scale factor $a$ is assumed, we used the flexibility of GalaxyNet to directly measure this scaling. Interestingly, we could confirm the linear relation for three of the parameters: the characteristic mass $M_1$ (where the efficiency peaks), and the slopes $\beta$ and $\gamma$. However, we found that the normalisation does not follow a linear relation with $a$ up to high redshift, but only up to $z\lesssim0.7$, and then becomes constant at some maximum efficiency. Moreover, while the scaling for $\beta$ and $M_1$ are linear, the evolution in $M_1$ is stronger and $\beta$ is shallower compared to \textsc{emerge}.

(iv) \textit{Large scale clustering}: Using GalaxyNet trained with reinforcement learning in combination with one of the largest cosmological simulations, the Huge MultiDark Planck simulation, we were able to predict galaxy clustering on very large scales as function of their stellar mass and SFR. We found that the peak of the BAO signal slightly depends on stellar mass at all redshifts, where massive galaxies peak at smaller scales. However, the zero-crossing of the correlation function happens at the same comoving scale of $123\Mpc\,h^{-1}$ independent of stellar mass and redshift. We further computed the galaxy bias as function of stellar mass and provided fitting functions that can be used to analytically compute large scale clustering and cosmic variance. Finally, at low redshift we observed GalaxyNet+RL to predict stronger clustering of passive galaxies for all but the most massive galaxies. However, for galaxies with $\log(m_*/\Msun) \gtrsim 11$ we found stronger clustering of active galaxies.

Overall, we conclude that training machine learning models such as wide \& deep neural networks with reinforcement directly on observed data rather than on data provided by models can be a very valuable approach for future studies of galaxy formation and cosmology. The advantage over typical empirical models is the increased flexibility, as it is not necessary to parameterise relations beforehand. The advantage over machine learning methods that employ supervised learning is that the model can be trained directly on observed data, even when key features such as the halo mass cannot be observed, so that no data produced by some other model with potential biases are needed. Instead, these models can be used to study the connection between galaxies and dark matter haloes without relying on any assumptions on baryonic physics.

In it current form, GalaxyNet can predict the stellar mass and SFR for any dark matter halo up to $z=8$ without any knowledge on its formation history other than the peak mass and corresponding scale factor and growth rate. This has the big advantage that only a few features need to be known and mock data can be created quickly and easily. However, the disadvantage is that the two target values are not modelled self-consistently, i.e. the stellar mass of a galaxy is not required to correspond to the integrated SFR. Therefore, future modelling endeavours will concentrate on predicting only the SFR and then computing the stellar mass from the star formation history. This approach firstly requires the input haloes to be linked through time, and secondly requires the time between snapshots to be sufficiently small, so that the time integral can be computed with adequate accuracy. This could be achieved by using halo merger trees as input data. However, simple feed-forward neural networks would not be the ideal method to deal with the data. Instead more advanced models, such as recurrent neural networks \citep[RNNs][]{Rumelhart:1986aa,Hochreiter:1997aa} could be applied.


\section*{Acknowledgements}

We are grateful to
Andrew Hearin,
Ben Hoyle,
Luisa Lucie-Smith,
Jochen Weller,
and Simon White
for useful discussions.
%
BPM and JAO acknowledge an Emmy Noether grant funded by the Deutsche Forschungsgemeinschaft (DFG, German Research Foundation) -- MO 2979/1-1.
This research was supported by the Excellence Cluster ORIGINS which is funded by the Deutsche Forschungsgemeinschaft (DFG, German Research Foundation) under Germany´s Excellence Strategy -- EXC-2094 -- 390783311, and by the DFG cluster of Excellence UNIVERSE.
%
The cosmological simulations used in this work were carried out at the Freya Cluster at the Max Planck Computing and Data Facility in Garching (MPCDF, mpcdf.mpg.de), and the GCS Supercomputer SuperMUC at Leibniz Supercomputing Centre (LRZ, lrz.de). We acknowledge the Gauss Centre for Supercomputing e.V. (gauss-centre.eu) and the Partnership for Advanced Supercomputing in Europe (PRACE, prace-ri.eu) for funding the MultiDark simulation project.
The CosmoSim database used in this paper is a service by the Leibniz-Institute for Astrophysics Potsdam (AIP).
The MultiDark database was developed in cooperation with the Spanish MultiDark Consolider Project CSD2009-00064.
%
We thank the developers of \texttt{Astropy} \citep{Astropy-Collaboration:2013aa,Astropy-Collaboration:2018aa}, \texttt{NumPy} \citep{van-der-Walt:2011aa}, \texttt{SciPy} \citep{Virtanen:2020aa}, \texttt{Jupyter} \citep{Ragan-Kelley:2014aa}, \texttt{Matplotlib} \citep{Hunter:2007aa}, \texttt{HaloTools} \citep{Hearin:2017aa}, \texttt{Scikit-learn} \citep{Pedregosa:2011aa}, \texttt{TensorFlow} \citep{Abadi:2015aa,Abadi:2016aa}, and \texttt{Keras} \citep{Chollet:2015aa} for their very useful free software. The Astrophysics Data Service (ADS) and \texttt{arXiv} preprint repository were used intensively in this work.




\bibliographystyle{mnras}
\bibliography{astro}







\bsp	
\label{lastpage}
\end{document}